\documentclass[usenatbib]{mn2e}

\newcommand{\msun}{{\it M}$_{\odot}$}

\newcommand{\etal}{{\it et al.}}
\newcommand{\ie}{{\it i.e.}}
\newcommand{\eg}{{\it e.g.}}
\newcommand{\be}{\begin{equation}}
\newcommand{\ee}{\end{equation}}
\newcommand{\bm}{\begin{displaymath}}
\newcommand{\eem}{\end{displaymath}}

\newcommand{\kmsMpc}{km~s$^{-1}$ Mpc$^{-1}$}

\newcommand{\mnras}{MNRAS}
\newcommand{\apj}{ApJ}
\newcommand{\aj}{AJ}
\newcommand{\pasp}{PASP}

\newcommand{\aap}{A\&A}
\newcommand{\apjs}{ApJS}
\newcommand{\apjl}{ApJL}
\newcommand{\araa}{ARA\&A}

\usepackage{epsfig}
\begin{document}

\title[Galaxy Zoo: Passive Red Spirals]{Galaxy Zoo: Passive Red Spirals$^*$. }
\author[K. L. Masters  \etal]{Karen L. Masters$^{1}$, Moein Mosleh$^{2,3}$, A. Kathy Romer$^2$, Robert C. Nichol$^1$, \newauthor Steven P. Bamford$^4$, Kevin Schawinski$^{5}$, Chris J. Lintott$^6$, Dan Andreescu$^7$, \newauthor  Heather C. Campbell$^{2,1}$,  Ben Crowcroft$^8$, Isabelle Doyle$^1$, Edward M. Edmondson$^1$, 
\newauthor Phil Murray$^{9}$,  M. Jordan Raddick$^{10}$, An\v{z}e Slosar$^{11}$,  
Alexander S. Szalay$^{10}$, \newauthor  and Jan Vandenberg$^{10}$\\
$^1$Institute for Cosmology and Gravitation, Dennis Sciama Building, University of Portsmouth,  Burnaby Road, Portsmouth, PO1 3FX, UK \\  
$^2$Astronomy Centre, University of Sussex, Falmer, Brighton BN1 9QJ, UK \\ 
$^3$Leiden Observatory, Leiden University, P.O. Box 9513, 2300 RA Leiden,The Netherlands\\ 
$^4$Centre for Astronomy \& Particle Theory, University of Nottingham, University Park, Nottingham, NG7 2RD, UK\\
$^{5}$Einstein Fellow/Yale Center for Astronomy and Astrophysics, Yale University, P.O. Box 208121, New Haven, CT 06520, USA \\
$^{6}$Oxford Astrophysics, Department of Physics, University of Oxford, Denys Wilkinson Building, Keble Road, Oxford, OX1 3RH, UK\\
$^{7}$LinkLab, 4506 Graystone Ave., Bronx, NY 10471, USA\\
$^8$Portsmouth Grammar School, High Street, Portsmouth, PO1 2LN, UK\\ 
$^{9}$Fingerprint Digital Media, 9 Victoria Close, Newtownards, Co. Down,Northern Ireland, BT23 7GY, UK\\
$^{10}$Department of Physics and Astronomy, The Johns Hopkins University, Homewood Campus, Baltimore, MD 21218, USA\\
$^{11}$Berkeley Center for Cosmo. Physics, Lawrence Berkeley National Lab. \& Physics Dept.,Univ. of California, Berkeley CA 94720, USA \\\ 
\\
$^*$This publication has been made possible by the participation of more than 160,000 volunteers in the Galaxy Zoo project. \\ Their contributions are individually acknowledged at \texttt{http://www.galaxyzoo.org/Volunteers.aspx}. \\
\\
{\tt E-mail: karen.masters@port.ac.uk}
}

\date{Accepted by MNRAS}
\pagerange{1--20} \pubyear{2010}

\maketitle

\begin{abstract}
 We study the spectroscopic properties and environments of red (or passive) spiral galaxies found by the Galaxy Zoo project. By carefully selecting face-on, disk dominated spirals we construct a sample of truly passive disks (\ie ~they are not dust reddened spirals, nor are they dominated by old stellar populations in a bulge). As such, our red spirals represent an interesting set of possible transition objects between normal blue spiral galaxies and red early types, making up $\sim6\%$ of late-type spirals. We use optical images and spectra from SDSS to investigate the physical processes which could have turned these objects red without disturbing their morphology. 
 We find red spirals preferentially in intermediate density regimes. However there are no obvious correlations between red spiral properties and environment suggesting that {\it environment alone is not sufficient to determine if a galaxy will become a red spiral}.  Red spirals are a very small fraction of all spirals at low masses ($M_\star < 10^{10}$\msun), but are a significant fraction of the spiral population at large stellar masses showing that {\it massive galaxies are red independent of morphology}.  We confirm that as expected, red spirals have older stellar populations and less recent star formation than the main spiral population.  While the presence of spiral arms suggests that major star formation cannot have ceased long ago (not more than a few Gyrs), we show that these are also not recent post-starburst objects (having had no significant star formation in the last Gyr), so {\it star formation must have ceased gradually}. Intriguingly, red spirals are roughly four times as likely than the normal spiral population to host optically identified Seyfert/LINER (at a given stellar mass and even accounting for low luminosity lines hidden by star formation), with most of the difference coming from objects with LINER-like emission. We also find a curiously large optical bar fraction in the red spirals ($70\pm5\%$ verses $27\pm5\%$ in blue spirals) suggesting that {\it the cessation of star formation and bar instabilities in spirals are strongly correlated}. 
 We conclude by discussing the possible origins of these red spirals. We suggest they may represent the very oldest spiral galaxies which have already used up their reserves of gas - probably aided by strangulation or starvation, and perhaps also by the effect of bar instabilities moving material around in the disk. We provide an online table listing our full sample of red spirals along with the normal/blue spirals used for comparison. 
\end{abstract}


\section{Introduction}
The advent of large galaxy surveys like the Sloan Digital Sky Survey (SDSS) in which photometry (and therefore colours) are readily available for millions of objects has lead to the common use of optical colours to define ``early" and``late" type galaxy samples \citep[e.g.][]{S09,S08,LP07,C07,B06,C05}. This method is particularly favoured since obtaining morphologies for large numbers of galaxies has until recently been impossible. This simplification is justified since it has been shown many times that the majority of galaxies follow a strict colour-morphology relation. For example \citet{M09} argued that 85\% of galaxies to z$\sim$1 are either red, bulge-dominated galaxies or blue, disk dominated galaxies; while \citet{C06} showed a similar result for 22000 low redshift galaxies (both using automated methods for morphological classification). 

However the clear correlation between colour and morphology is surprising, given that the colours of galaxies are determined primarily by their stellar content (and therefore their recent star formation history, mostly within the last Gyr) while the morphology is primarily driven by the dynamical history. The clear link between colour and morphology then gives a strong indication that the timescales and processes which drive morphological transformation and the cessation of star formation are strongly related - at least in most cases. In this paper however, we consider a class of object (the red spirals) where the link described above appears to be broken.

 Since the morphology-density relation was first quantified \citep{D80} many mechanisms have been proposed for the transformation of blue, star forming, disk galaxies in low density regions of the universe, to red, passive, early type galaxies in clusters. A recent review of many of the proposed mechanisms, and the evidence supporting them, can be found in \citet{BG06}. Clearly two things must happen for a star forming blue spiral galaxy to turn into a passive red early type. First, star formation must cease (which can indirectly alter the morphology by causing spiral arms and the disk in general to fade, possibly producing an S0 or lenticular from a spiral), and secondly, in order to produce a {\it bona fide} elliptical the same, or a different process must also dynamically alter the stellar kinematics of the galaxy. 
 
The presence of an unusually red or passive (\ie. non-star forming) population of spiral galaxies in clusters of galaxies was first noticed by \citet{V76} in the Virgo cluster. Later studies of distant cluster galaxies in {\it Hubble Space Telescope} (HST) imaging also revealed a significant number of so-called ``passive'' spiral galaxies with a lack of on-going star formation (Couch et al. 1998; Dressler et al. 1999; Poggianti et al. 1999).  Passive late-type galaxies were identified at lower redshifts in the outskirts of SDSS clusters by \citet{G03}, using concentration as a proxy for morphology. Passive spirals in a cluster at $z\sim0.4$ were studied by \citet{M06} who found star formation histories from GALEX observations consistent with the shutting down of star formation from strangulation (as described by \citealt{Bekki02}). Passive spirals have also been revealed in a cluster at $z\sim 0.1$ in the STAGES survey, using HST morphologies \cite{W09}, rest frame NUV-optical SEDs \citep{W05} and 24$\mu$m data from Spitzer \citep{G09}. In that series of papers, ``dusty red late-types" and ``optically passive late-types" are found to be largely the same thing, with a non-zero (but significantly lowered) star formation rate revealed by the IR data. 

Red spirals/late types have been studied in several recent papers \citep{Lee08,D09,HC09,CH09}, as well as \citet{MR09} who talk about blue passive galaxies (\ie ~ galaxies with blue colours, but showing no indication of recent star formation in their spectra) which mostly appear to have late type morphologies and have very recently shut down star formation. These might be the progenitors of the red spirals. \citet{Bundy09} have studied the redshift evolution of red sequence galaxies with disk like components in COSMOS and use it to estimate that as many as 60\% of spiral galaxies must pass through this phase on the way to the red sequence - making it an important evolutionary step. 

The Galaxy Zoo project \citep{L08} revealed the presence of a significant number of visually classified spiral galaxies which are redder than the blue cloud (between 16-28\% of the total galaxy population depending on environment, \citealt{B09}). In this paper we study in more detail the physical properties and environments of this population of red spiral galaxies. Galaxies drawn from this population have the morphological appearance of spiral galaxies with a distinct spiral arm structure, but have rest-frame colours which are as red as a typical elliptical galaxy, indicating little or no recent star formation activity. We are studying these objects in order to identify the physical process which is most important in their formation. 

 It is clear that all spiral galaxies can be affected by various physical processes as they evolve -- in this paper we attempt to identify which are most important for red spirals, asking how they are able to shut down star formation while retaining their spiral morphology. A list of possible mechanisms includes processes that depend on environment such as: (1) galaxy-galaxy interactions: in high density regions there is an increased probability of interaction with other galaxies. Most major mergers destroy spiral structure \citep{TT72} unless they involve very gas rich progenitors \citep{H09}, but some interactions can be quite gentle (\eg~ Walker, Mihos \& Hernquist 1996), for example minor-mergers, tidal interactions etc. (2) Interaction with the cluster itself also occurs and can remove the gas which forms the reservoir for star formation. This can be due to tidal effects \citep[e.g.][]{Gn03}, or interaction with the hot intercluster gas, either through thermal evaporation \citep{CS77} or ram pressure stripping \citep{GG72}. (3) Processes like harrassment \citep{M99} and starvation or strangulation \citep{L80,Bekki02} have also been shown to have a significant effect on late type galaxies. Harrassment refers to the heating of gas by many small interactions, while starvation or strangulation refers to the gradual exhaustion of disk gas after the hot halo has been stripped away. These mechanisms both occur at much larger cluster radii (\ie ~lower densities) than the "classic" environmental effects. Internal mechanisms could be more important.  For example, (4) the latest semi-analytical models of galaxy formation all invoke feedback from a central massive black hole (or active galactic nuclei; AGN) to explain the most massive red elliptical galaxies \citep{G04,S05,Sc06,Cr06,Bower06}, although the effect of this process on disk galaxies has been studied less, it still may have some effect (Okamoto, Nemmen \& Bower 2008). (5) Another culprit could be bar instabilities in spiral galaxies which drive gas inwards (eg. Combes \& Sanders 1981) and may trigger AGN activity and/or central star formation (eg. Shlosman et al. 2000), perhaps using up the reservoir of gas in the outer disk and making spirals red. (6) Finally red spirals could simply be old spirals which have used up all their gas in normal star formation activities without having any major interactions. In normal spirals, the gas that feeds ongoing star formation comes from infall of matter from a reservoir in the outer halo \citep{BG06}. As first suggested by Larson, Tinsley \& Caldwell (1980) and expanded by \citet{Bekki02} the removal of gas from this outer halo (``strangulation", or ``starvation") will cause a gradual cessation of star formation proceeding over several Gyrs.

We describe the sample and data along with the selection of the ``red spirals"  in more detail in Section 2. In Section 3 we discuss the stellar populations and star formation history of the red spirals. In Section 4 we discuss their environment and the environmental dependence of star formation. The impact of AGN is considered in Section 5, and bar instabilities are discussed in Section 6. In Section 7 we discuss the plausible mechanisms for formation of the red spirals and future directions which could be taken to distinguish between them. We present a summary and conclusions in Section 8. The adopted cosmological parameters throughout this paper are $\Omega_{m}$ = 0.3, $\Omega_{\Lambda}$ = 0.7 and $H_{0} = 70$ \kmsMpc.

\section{Sample Selection and Data}

The sample of visually classified spiral galaxies used in this paper is drawn from the Galaxy Zoo (GZ1) clean catalogue (Lintott et al. 2008) which, by an order of magnitude, is the largest morphologically classified sample of galaxies.  To make this unprecedented sample, over 160,000 volunteers visually inspected images of SDSS galaxies independently via an internet tool (the original GZ1 sample was selected from the SDSS Data Release 6; \citealt{A-M08}). For more information on the classification process in GZ1 and the conversion of multiple classifications per galaxy to redshift bias corrected type ``likelihoods" see \citet{L08} and \citet{B09}. In brief, we use the spiral likelihood $p_{\rm spiral}$ and (in order to tell if the spirals have visible spiral arms) the quantities $p_{\rm CW}$ and $p_{\rm ACW}$ which describe how likely the spiral is to have ``clockwise" or ``anti-clockwise" arms respectively. 

We make a volume-limited sample of galaxies from the GZ1 catalogue by selecting only objects from the SDSS Main Galaxy Sample \citep{S02} with spectroscopic redshift between 0.03 $<$ z $<$ 0.085 and by limiting the sample to an absolute magnitude of $M_{r}$ $<$ $-20.17$. This redshift range is picked to remove problems with peculiar velocities at the low redshift end (which would bias distance dependent quantities) and the upper limit is chosen as a compromise between sample size and luminosity range, (and also a redshift up to which reliable local densities - see below - are available).

Our photometric quantities are taken from the SDSS DR6 \citet{A-M08}. We use model magnitudes for colours, and Petrosian magnitudes for total luminosities. We also make use of shape/structural parameters from SDSS, namely the axial ratio ($a/b$) from the $r$-band isophotal measurement which is used as a proxy for disk inclination; and {\tt fracdeV} (or $f_{\rm DeV}$) - the fraction of the best fit light profile which comes from the de Vaucouleurs fit (as opposed to the exponential fit) and which is used as a proxy for bulge size in these visually classified spirals (as discussed in \citealt{Ma09}).

SDSS 3'' fibre spectra are available for all galaxies in our sample. We use equivalent widths (EW) and absorption line indices measured for SDSS DR7 spectra by the MPA-Garching group whose methods are described in \citet{T04}\footnote{http://www.mpa-garching.mpg.de/SDSS/}.  

 As an estimate of stellar mass, we use the empirical fit in \citet{Baldry06} which gives a stellar mass-light ratio in the $r$-band as a function of the $(u-r)$ colour of the galaxy  (based on calculations of stellar masses from \citet{K03} and \citet{G04} using the \citet{BC03} stellar population models). Obviously this is an oversimplification of the calculation of stellar masses, not taking into account the varied star formation histories of galaxies with the same colour and luminosity; however the method does have the advantage of being simply related to only two measured physical properties of the galaxies (colour and luminosity). 

The local galaxy densities used in this paper are identical to those used in Bamford et al. (2009). The details of the method are described in \citet{Baldry06}. Briefly, $\Sigma_{N}$ is determined by $ N/(\pi$ $d_{N}^{2})$, where $d_{N}$ is the projected distance to the $N$th nearest galaxy (with $M_r<-20$) within $cz \pm$1000 km s$^{-1}$. The final value used, $\Sigma$, is the average of $\Sigma_{N}$ for N = 4 and 5. These local densities include a correction for redshift incompleteness due to fiber collisions by considering the photometric redshift likelihood distributions for galaxies without spectra. Typical values of $\Sigma$ range from 0.05 Mpc$^{-2}$ in voids to 20 Mpc$^{-2}$ in clusters \citep{Baldry06}. 

\subsection{Selection of Red Spiral Galaxies}

We wish to select from the GZ1 volume limited sample a subset of truly passive red disk-dominated spiral galaxies (\ie~ not dust reddened, nor red because they are dominated by bulge emission). We use a cut in the spiral likelihood (corrected for the small bias described in the Appendix of \citealt{B09}) of $p_{\rm spiral} \geq 0.8$. Two things can complicate the picture in a sample selected purely by colour and spiral likelihood:
\begin{enumerate}
\item Dust reddening of edge-on spiral galaxies
\item Contamination by early-type spirals and/or S0s.
\end{enumerate}
\citet{Ma09} show that the impact of dust reddening can be significant for inclined spiral galaxies, and that spiral galaxies with large bulges (as measured by {\tt fracdev}) are intrinsically red. Therefore with no cuts on inclination or bulge size, a red spiral sample will be dominated by inclined dust reddened spirals, and spirals with large bulges. We choose in this work to study only the most face-on spiral galaxies, requiring $\log (a/b) < 0.2$. This will minimize the impact of dust reddening on the sample (although we note that even at face-on spirals can be dust reddened; for example \citealt{Ma09} find a median Balmer decrement of $0.3\pm0.3$ mags in the centres of face-on GZ1 spirals). 

\begin{figure*}
\includegraphics{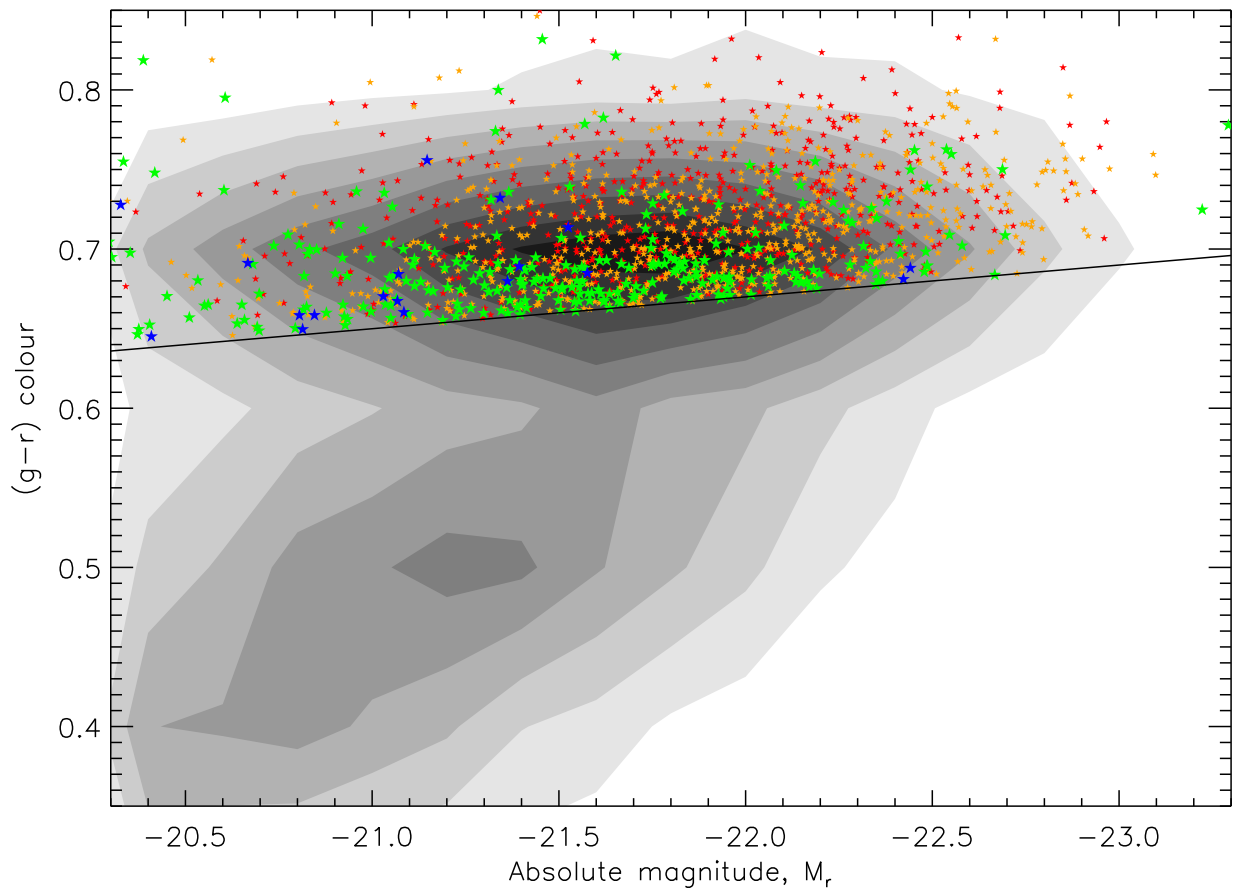}
\caption{Colour magnitude diagram of the``face-on" ($\log (a/b)<0.2$) volume limited clean GZ1 sample. The greyscale contours show the location of galaxies in the red sequence and blue cloud. The solid line indicates the blue edge of the red sequence of GZ1 early types. All red spirals (having visible spiral arms) are shown colour coded by {\tt fracdev} (blue: $f_{\rm DeV}<0.1$, which roughly corresponds to type Sd; green: $0.1<f_{\rm DeV}<0.5$, or Sb-Sc; orange: $0.5<f_{\rm DeV}<0.9$, or Sa-S0/a; red: $f_{\rm DeV}>0.9$ which are spirals having very large bulges - but note that spiral arms are still visible here, so these are not S0s.). Our final red spiral sample selects only those galaxies with $f_{DeV}<0.5$ (\ie~ the blue and green points) to select against spirals dominated by light from the bulge. \label{cmdiagram}}
\end{figure*}

Both \citet{L08} and \citet{B09} show that contamination of S0s into the GZ1 clean spiral sample should be small - an estimate of 3\% contamination is made.  Furthermore we use {\tt fracDev} $ \leq 0.5$ to select ``disky" spirals. As illustrated in \citet{Ma09} using a small sample of GZ1 spirals with B/T from the Millennium Galaxy Catalogue \citep{L03} this selects spiral galaxies with $B/T<0.25$, or of types $\sim$Sb--Sc (where this relation between typical $B/T$ and Hubble type comes from \citealt{SdV86}). One caveat to note here is that both {\tt fracdeV} and $B/T$ are light-weighted quantities, so in using the same {\tt fracdeV} limit for both red and blue spirals will results in the red spirals having a slightly {\it lower} upper limit in bulge {\it mass} fraction than the blue spirals. This slight bias should only make the conclusions below stronger, as the red spirals may on average be more disk dominated than the comparison blue spiral sample.

Finally we require that spiral arms be visible to GZ1 users, using $p_{\rm CW} > 0.8$ or $p_{\rm ACW} > 0.8$. The presence of spiral arms gives an indication that these red spirals may have only recently stopped forming stars, since spiral structures are expected to persist for only a short time after star formation ceases (for example \citet{Bekki02} found that spirals arms persisted for only a few Gyrs after the gas which provides the reservoir for ongoing star formation was removed). 

We show in Figure \ref{cmdiagram}, the colour magitude diagram of the ``face-on" volume limited clean GZ1 sample. The locus of the galaxy population is illustrated by the greyscale contours, and the positions of visually classified spirals in the red sequence are hi-lighted. The best fit to the red sequence of early types ($p_{\rm el}\geq0.8$) is $(g-r) = 0.73 - 0.02 (M_r + 20)$, and the scatter is $\sigma = 0.1$ mags. We therefore define the blue edge of the red sequence as \be (g-r) = 0.63 - 0.02 (M_r + 20), \ee  which is indicated by the solid line in Figure \ref{cmdiagram}, and define a red spiral as one redder than this limit. Since the limit depends on magnitude, we point out that in our definition a blue spiral with a high luminosity could actually be slightly redder than a red spiral with a low luminosity. 

\begin{figure*}
\includegraphics[width=160mm]{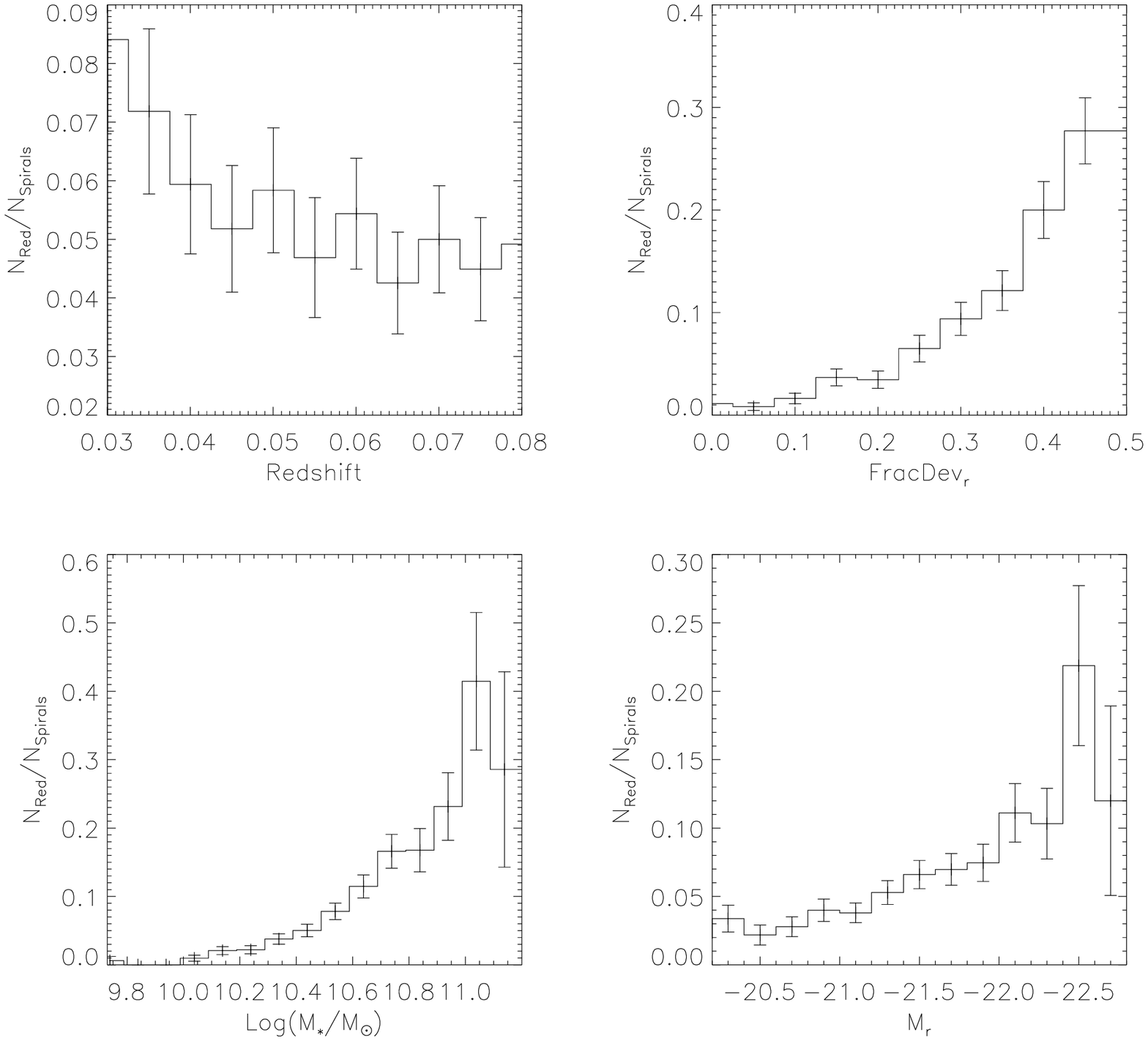}
\caption{This figure shows the fraction of all face-on, disky GZ1 spirals which are found to be red (\ie~ $N_{\rm red}/N_{\rm spiral}$) as a function of (1) redshift, (2) {\tt fracdeV} (or bulge size), (3) stellar mass, and (4) $r$-band absolute magnitude. $\sqrt{N}$ counting errors are shown. \label{comparison}}
\end{figure*}

The colour cut used here to select red spirals is different from than that used by either \citet{B09}, who used a cut in $(u-r)$ vs. stellar mass; or \citet{Sk09}, who used k-corrected (to $z=0.1$) magnitudes, which reddens the red sequence by $\sim 0.1$ mags. Our cut is more conservative that either of those papers; they effectively select red spirals as being redder than most of the blue cloud, while we select red spirals to be {\it as red as most elliptical galaxies}. The difference in the two approaches is especially large at the low luminosity/stellar mass end of the galaxy population where (as can be seen in Figure 1) the blue cloud is significantly bluer than the red sequence. However, the major difference in our sample and all previous ones used to study passive or red spirals is that we have made an effort to remove dust reddened inclined spirals, which has not been done before \citep[e.g.][]{B09,Sk09,W09}.

Our final sample consists of 5433 face-on disky spirals with visible spiral arms, 294 of which (or 6\%) are classified as red. In what follows, we refer to the remaining 5139 spirals as our comparison sample a refer to them as either ``blue" spirals, or ``normal" spirals. We provide an electronic table listing the SDSS names, positions and optical photometry parameters used in this paper for these two samples (our red spirals and comparison sample). A sample of this table is provided in the Appendix. 

Figure \ref{comparison} shows the redshift, {\tt fracdeV} (for bulge size), luminosity and stellar mass distributions of the red spirals compared to all spirals (in the face-on, disky sample). Red spirals are more likely to be found at the higher luminosity and larger {\tt fracdeV} (and hence larger bulge) end of the spiral distribution showing that as is well known, luminous spirals with large bulges (earlier spiral types) are more likely to be red. There is a slight trend with redshift which can be explained by effect of the luminosity distribution in a volume limited sample. 

 As also discovered by \citet{W09}, red (or passive) spirals are an insignificant fraction of the total spiral population at stellar masses below $\sim 10^{10}$ \msun~(although note that \citealt{B09} shows that in the densest regions the few remaining low mass spirals are mostly red) but are a significant fraction of spirals at large stellar masses. This shows that {\it massive galaxies are red independent of morphology}.  However we note here that this mass distribution may be slightly biased by our decision to classify spirals as red only if they are {\it as red as ellipticals}. This cut biases against low mass red spirals as the blue cloud at low masses/luminosities moves further from the red sequence. Our sample is also incomplete for red objects below $10^{10.3}$\msun~ which corresponds to the luminosity limit for the reddest objects. Finally, we comment that the stellar population models of \citet{BC03} on which the \citet{Baldry06} stellar masses are based, struggle to make red low mass objects (since all their red objects must be very old, and therefore have very high mass-light ratios) so the use of this stellar mass estimate may anyway bias high the masses of the redder spirals. A more detailed modelling of the star formation history using stellar populations which account for the known populations of young red stars \citep[the thermally pulsing asymptotic giant branch (TP-AGB) phase included in][]{M05} would provide more reliable stellar masses for these objects.

\begin{figure*}
\includegraphics[width=85mm,angle=-90]{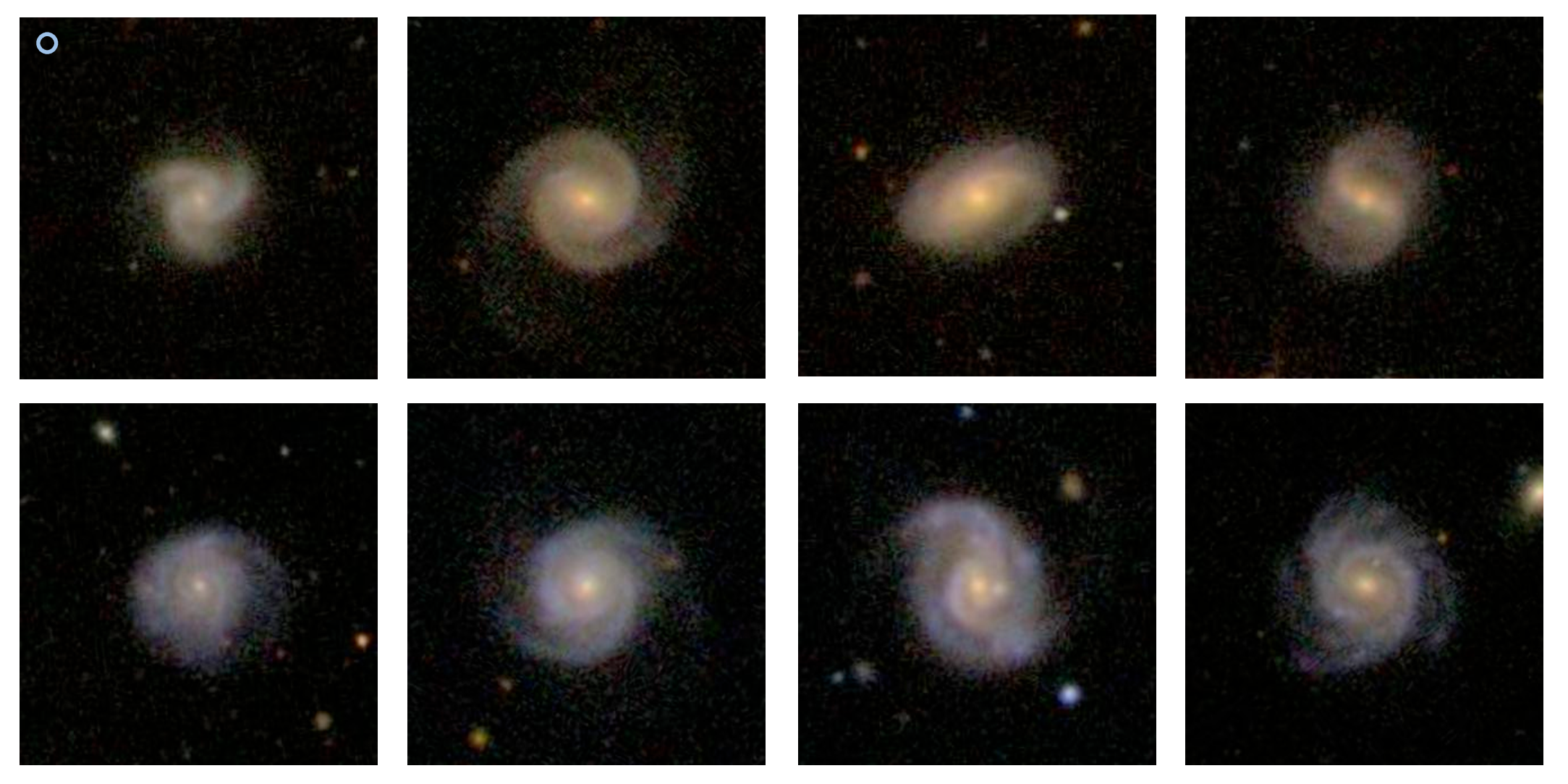}
\caption{{\it Top row: } examples of face-on, red, disky spirals with visible arms (as described in the text), {\it Bottom row}: face-on, blue, disky spirals picked to have similar redshifts, absolute $r$-band magnitudes, angular sizes and {\tt fracdeV} values as the red spiral immediately above them. Images are SDSS $gri$ composites, all $1\arcmin \times 1\arcmin$ in size (these are some of the largest angular size galaxies in our sample). The size of the SDSS fibre is indicated at top left of the top left image. Galaxies are (in red/blue pairs from left to right): SDSS J131428.83+334109.2 and SDSS J130058.63+395132.1 at $z\sim0.04$, $M_r\sim-20.3$; SDSS J082959.05+304340.1 and SDSS J100515.00+513545.9 at $z\sim0.05$ and $M_r\sim-21.7$; SDSS J104034.48+004902.6 and SDSS J123317.60+575620.1 at $z\sim 0.07$ and $M_r\sim -22.6$; and SDSS J134248.47+145553.5 and  SDSS J155357.57+383923.9 at $z\sim0.08$ and $M_r\sim-22.3$. \label{red}}
\end{figure*}

\begin{figure*}
\includegraphics[width=68mm,angle=-90]{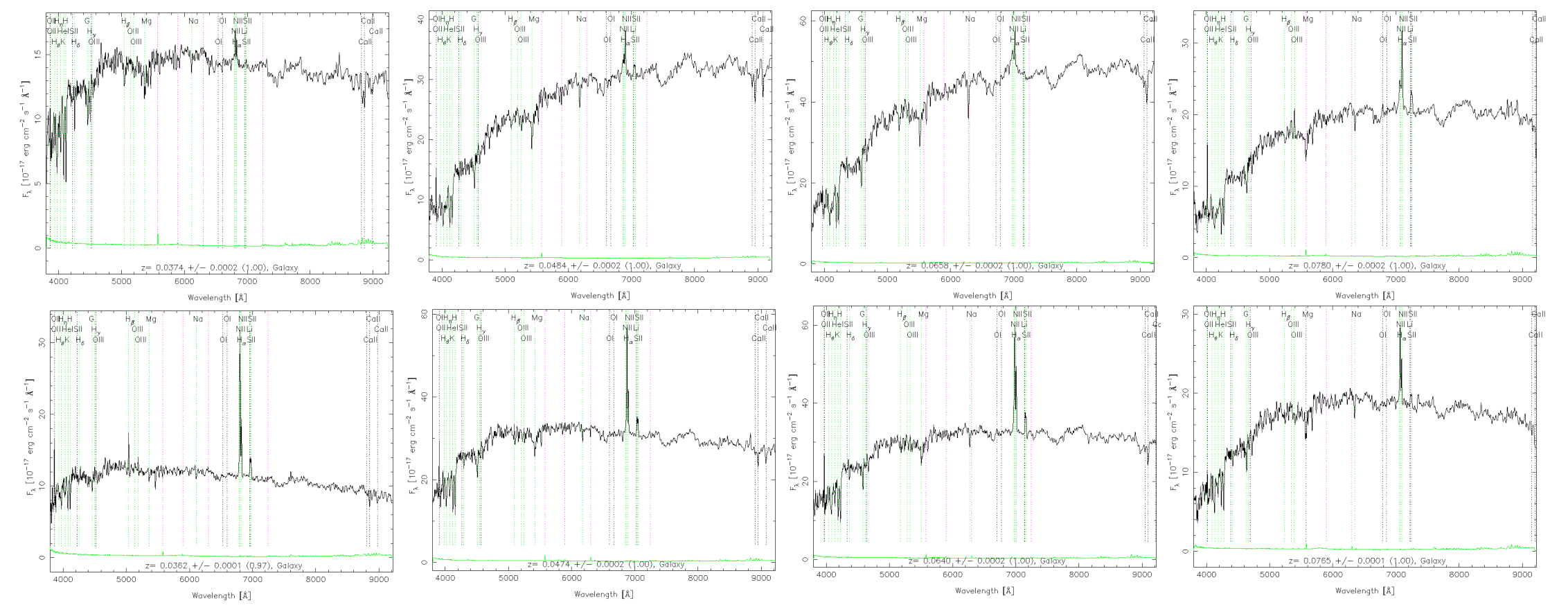}
\caption{SDSS fibre spectra of the spirals shown in Figure \ref{red}. Even in the central 3\arcsec, these spirals are clearly different in colour; the red spirals (top) mostly have continuum emission which increases towards long wavelengths (\ie ~ a red colour), while the blue spirals have continuum levels decreasing in that direction. Also note the significantly larger H$\alpha$ emission in the spectra of the blue spirals (indicative of more ongoing star formation), and the larger break at $\sim 4000 \AA$ in the spectra of the red spirals (indicating an older stellar population). \label{redspectra}}
\end{figure*}

Figure \ref{red} shows example images of red and blue spirals from our sample of face-on disky spirals. The blue spirals shown have been picked to have similar redshifts and absolute magnitudes as the red spiral directly above them. A range of redshifts are shown. In these $gri$ colour composite images the colour difference between red and blue spirals is quite clear to the eye. Figure \ref{redspectra} shows SDSS fibre spectra for the same galaxies in Figure \ref{red} (arranged in the same  order), and even in this small sample it is clear that red spirals show less (but not zero) nebular emission from on-going star formation than the blue spirals (since they have smaller H$\alpha$ emission lines) and have older stellar populations (the break at $\sim 4000\AA$ is larger in the red spirals than the blue spirals). 
 
\section{Star Formation in Red Spirals}
In this section, we use the SDSS fibre spectra to study in more detail the star formation history and mean stellar age of ``red spiral" galaxies with reference to our comparison sample of blue spirals.  Spectroscopic parameters are ideal for studying the stellar content of galaxies. A full star formation history model fit to the galaxy spectra is beyond the scope of this paper, but absorption line indices such as H$\delta_{A}$ and the 4000\AA {} break strength provide information about stellar content and recent star formation history of a galaxy, while $H\alpha$ and [OII] emission lines can indicate the presence of recent star formation. 

\subsection{Dust Content}
Before proceeding to use the spectral information to study the stellar content of the red spirals we first want to check the dust content of the red spirals relative to the blue spirals. It is possible that the red colours of the spirals could be due to an enhanced amount of extinction and reddening from dust (even in our specially selected sample of face-on spirals designed to minimize dust effects) rather than an ageing stellar population from a lack of recent star formation. In fact \citet{W09} find that ongoing star formation in their red spirals is obscured, although we note again that their objects include dust reddened inclined spirals which we have removed. 

 In order to quantify the levels of extinction we compare the difference in flux between the first two lines in the Balmer series (H$\alpha$ and H$\beta$). The expected flux ratio between these lines is $\Re_{\rm int} = F_\alpha/F_\beta = 2.76$, and has only a mild dependence on the temperature of the gas emitting the radiation (from $\Re_{\rm int} = 3.30$ at $2500^\circ$K to $\Re_{\rm int} =  2.76$ at $20000^\circ$K; \citealt{O89}). Therefore, deviations from the expected ratio can be used to measure the relative extinction between $H_\alpha$ (at 6562.8 $\AA$ in the $r$-band) and $H_\beta$ (at 4861.3 $\AA$ in $g$-band). The relative extinction between these two wavelengths in magnitudes is
\be
E(H_\beta - H_\alpha) = 2.5 \log \frac{\Re_{\rm obs}}{\Re_{\rm int}},
\ee
which is equivalent to a colour in narrow band filters at these wavelengths. 

\begin{figure}
\includegraphics[width=3.4in]{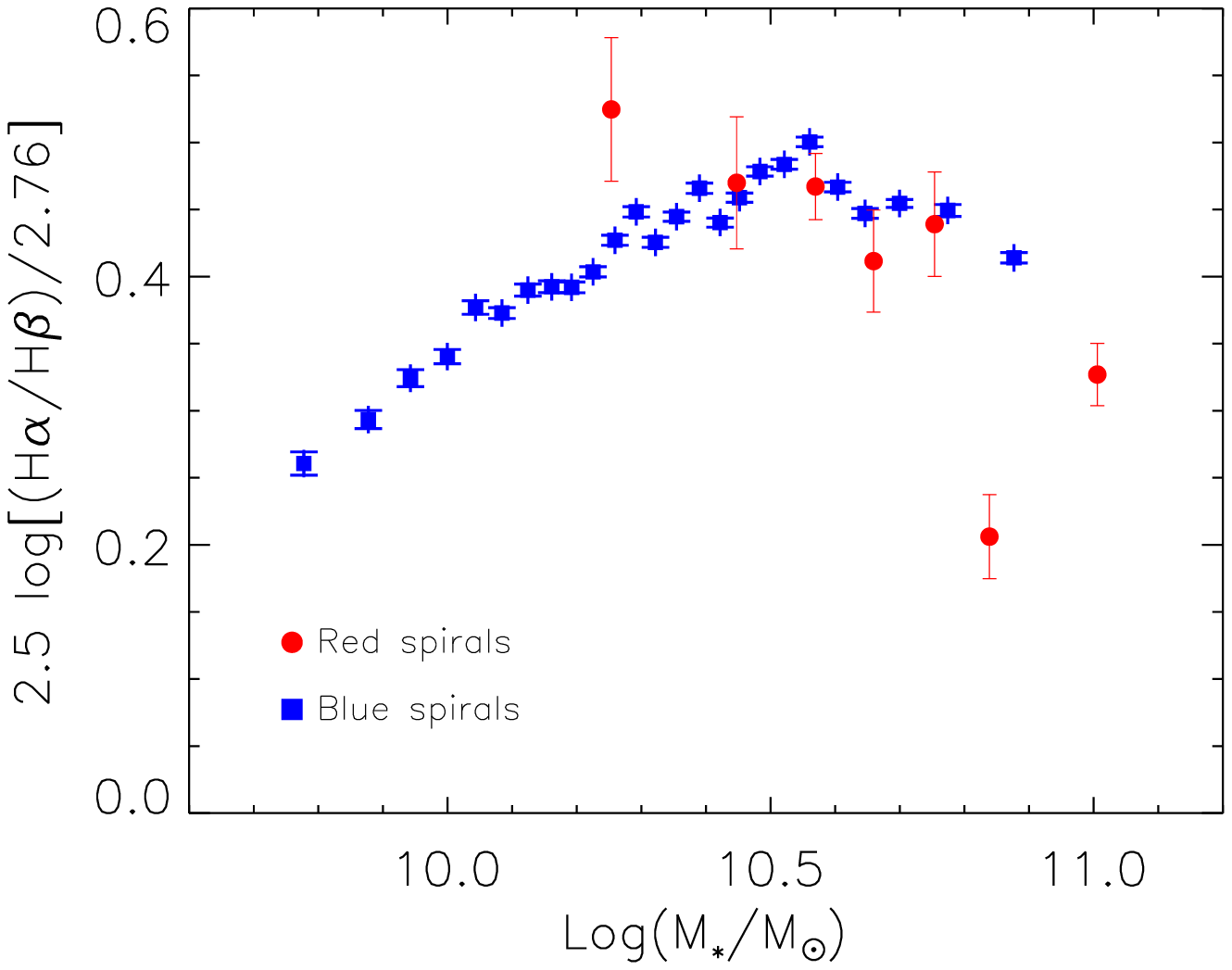}
\caption{Balmer decrement in the SDSS fibre (expressed in magnitudes of extinction between H$\alpha$ and H$\beta$) as a function of stellar mass for the blue (blue squares) and red (red dots) face-on disky spirals. Error bars indicate our estimate of the $1\sigma$ error on the median value.
\label{balmer}}
\end{figure}

In Figure \ref{balmer}, we plot the Balmer decrements from the 3'' SDSS fibre of blue and red spirals as a function of their stellar mass. This figure clearly indicates an increase in dust content of blue spirals as a function of stellar mass to a maximum at around $M \sim 10^{10.5}$\msun. The trend is less clear in the red spirals where the numbers are significantly smaller, but tentatively supports a drop off in dust content for the highest mass objects (as also seen in \citealt{Ma09}).  A negligible number of face-on spirals (either red or blue) have zero Balmer decrements in their central regions, showing that dust reddening always plays some role. However, we see no evidence for red spirals having significantly larger dust content than blue spirals at a given stellar mass (except perhaps in the lowest mass bin of the red spirals) and thus argue that the difference in colour is not due to differences in the dust content of the two populations.

\subsection{Recent Star Formation History}

Following the method in \citet{K03}, we use the Balmer absorption-line index H$\delta_{A}$ and the $D_{n}$(4000) break to estimate mean stellar ages and recent star-burst activity of our sample. 

The observed spectrum of a galaxy is the combination of flux from many stars. In galaxies with old stellar populations there are a large number of stellar absorption lines (mostly from ionized metals) which crowd together at around 4000\AA. The 4000\AA~ break which they produce is the strongest discontinuity seen in the optical spectrum of a galaxy. In galaxies with a younger stellar population, there are many more hot stars, and the metals in them are multiply ionized. This significantly reduces the strength of the break, making it a good indicator of the mean age of a galaxy's stellar population. We follow \citet{K03} and represent the strength of the 4000\AA~ break by the $D_{n}$(4000) index \citep{B99}.

H$\delta_{A}$ absorption lines can be used to trace the recent ($\leq 1$ Gyr) star formation activity of galaxies \citep{K03}.  H$\delta_{A}$ in the integrated galaxy spectra is almost entirely the contribution of A-type stars. The peak of this absorption feature thus occurs roughly 1 Gyr after a burst of star formation, once O and B stars have expired and the integrated light is dominated by A stars. 

\begin{figure}
\includegraphics[width=3.4in]{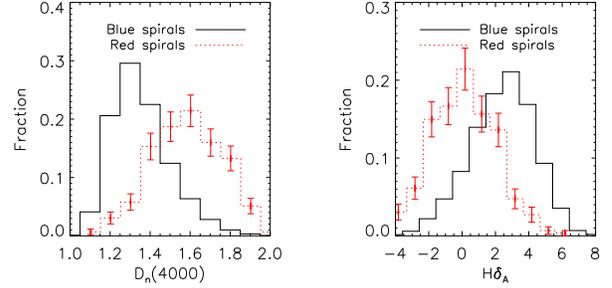}
\caption{Histograms of the distribution of H$\delta_{A}$ and $D_{n}$(4000) for the red and blue spiral samples (solid and dotted lines respectively). The y-axis shows the number in a given bin, relative to the total number in the respective sample. Counting errors are shown on the histogram for red spirals; due to the larger numbers of blue spirals the errors on those histograms are significantly smaller. 
\label{fig3}}
\end{figure}

\begin{figure*}
\includegraphics[width=6.8in]{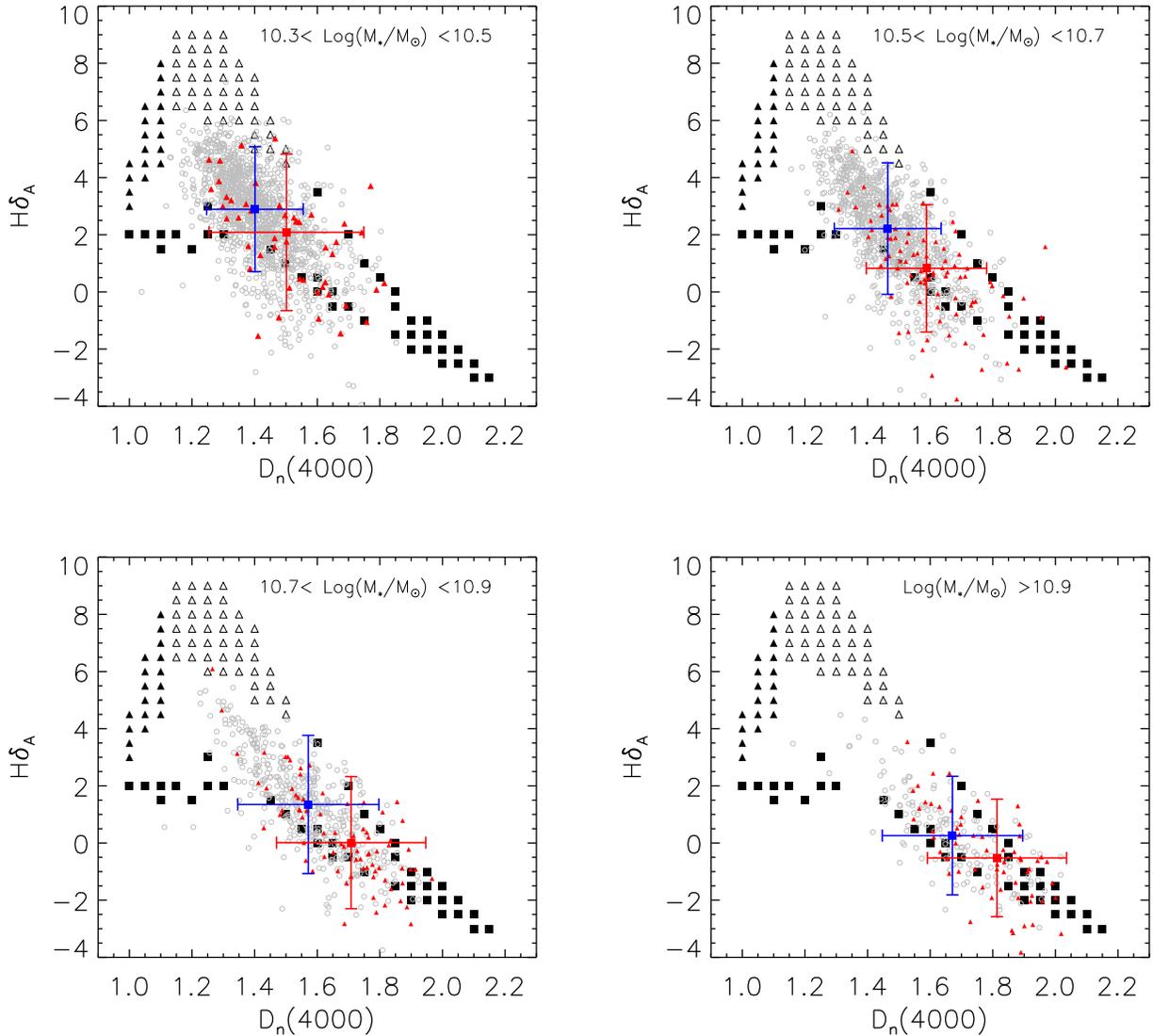}
\caption{H$\delta_{A}$ versus $D_{n}$(4000) is plotted for our sample of blue and red face-on, disky spirals. In each panel the blue spirals are shown by grey circles, and the red spirals are shown by red triangles. The red and blue error bars indicate the interquartile ranges for the red and blue spirals respectively. The different panels show the evolution with stellar mass, with the lowest mass galaxies at the upper left, down to the highest masses at the lower right. The red spiral galaxies on average show older stellar populations and less recent bursts of star formation in every mass bin. The data are plotted over galaxy models of \citet{K03} showing regions covered by their starbursting (filled triangle), post starburst (open triangle) and quiescent (filled squares) galaxies. 
\label{fig2}}
\end{figure*}

Histograms of the distribution of the two measurements for all the red and blue spirals are shown in Figure \ref{fig3}.  In Figure \ref{fig2}, we plot H$\delta_{A}$ versus $D_{n}$(4000) for our blue and red spirals samples split into 4 bins of stellar mass (starting at $\log (M_\star/M_\odot) = 10.3$ where the sample is complete for both red and blue spirals). These figures illustrate that the red spirals, on average (luminosity weighted), have both older stellar populations and are less likely to have had significant bursts of star formation in the last Gyr (they have larger mean $D_n(4000)$ and smaller H$\delta_A$).  The data are overplotted on the models of \citet{K03} showing the regions covered by their starburst (filled triangle), poststarburst (open triangle) and quiescent galaxies (filled squares). We can clearly see that red spirals at all masses are unlikely to fall in the part of the $D_n(4000)$-H$\delta_A$ plane which is occupied by star bursting or post star bursting galaxies (\ie. they are not above the main locus of points), but that a significant fraction do lie in the part of the plane in which galaxies are expected to have formed an insignificant fraction of their stars in the last 2 Gyrs \citep{K03}, as well as in the mid-region where galaxies are expected to have mixed star formation histories (ie. a mix of recent and much older star formation). In contrast, many of the blue spirals (especially at the lower stellar masses) are in the region of the plane occupied by galaxies with mixed star formation histories, although there are also star-burst/post star-burst as well as more quiescent blue spirals. 

 Figure \ref{fig2} illustrates clearly the mass dependence of spiral star formation history; lower mass spirals are more likely to have had recent star formation. This is true both for red and blue spirals in our sample, however in any given mass bin the red spirals have both lower H$\delta_A$ and larger $D_n(4000)$ than equivalent blue spirals. It is clearly not stellar mass alone which is determining the SF history of the red spirals. 

The range of stellar ages of red spirals are similar to early-type galaxies  (at least to first order considering the range of $D_n(4000)$ -- this does not account for metallicity effects which could cause a systematic offset) and they have not experienced recent burst of stars (in the last 1-2 Gyr); i.e. they are not post-starburst galaxies. However, the presence of spiral arms gives an indication that these red spirals must have recently been forming some stars, since spiral structures are expected to persist for only a short time after star formation completely ceases. The prevailing model for the origin of spiral structure in disk galaxies is the density wave theory proposed originally by \citet{LS64}. Discussing observations available at the time, which showed that galaxies without significant amount of interstellar gas do not have prominent spiral patterns, \citet{LS64} suggest that any spiral structure, even if present in the old stellar population would not be visible, due to a lack of gas and star formation.  More recent numerical simulations support this early picture of a relatively rapid fading of spiral arms after star formation ceases. For example \citet{Bekki02} found that spiral arms persisted for only a few Gyrs after the gas which provides the reservoir for ongoing star formation was removed. Observationally \citet{I07} use spatially resolved spectroscopy of the passive spiral SDSS J074452.52+373852.7 to show that this galaxy probably stopped forming stars about 1-2 Gyrs ago  -  and it's spiral structure is still just visible. Follow-up integral field unit, and/or higher S/N spectroscopy for a representative sample of the red spirals could measure population ages and test this picture further. Such work is planned.

 In the red spirals, current star formation is significantly reduced compared to the normal/blue spirals, {\it even at fixed stellar mass} but it does not appear to have completely stopped in all objects (as we will see below). Major amount of star formation cannot have ceased abruptly recently (as they are not post-starburst), or very long ago (more than a few Gyrs because of the presence of spiral arms). It is therefore possible to conclude that star formation either ceased abruptly in the red spirals between the post-starburst and spiral arm fading time scales (about one to a few Gyrs ago) or that there has been a gradual cessation of star formation in the red spirals over the past few Gyrs.

\section[]{The Effect of Environment}

The morphology-density relation \citep{D80} revealed that spiral galaxies favour lower density environments than early-types. 
However, recent work \citep[e.g.]{Ball08,B09,Sk09,D09} has shown that the colour-density relation is stronger than the morphology-density relation (ie. at a fixed morphology galaxies in higher density environments are redder, but at a fixed colour, there is little morphology-density dependence). We have already seen \citep{B09,Sk09} that GZ1 red spirals are more common in high density regions (and conversely blue early-types are more common in low density regions; \citealt{Sc09a}). There are also clear suggestions in both \citet{B09} and \citet{Sk09} that the peak of the red spiral distribution occurs in intermediate density regions. Therefore environment is clearly a candidate for driving the process which turns spirals red. As discussed in Section 1, various environmental mechanisms have been proposed to suppress star formation and turn galaxies red in high density regions. In this section we will consider the effect of environment on the red spirals in particular, looking at their preferred locations, and tracers of their star formation as a function of local density.

\subsection{The Environments of Red Spirals}
\begin{figure}
\includegraphics[width=84mm]{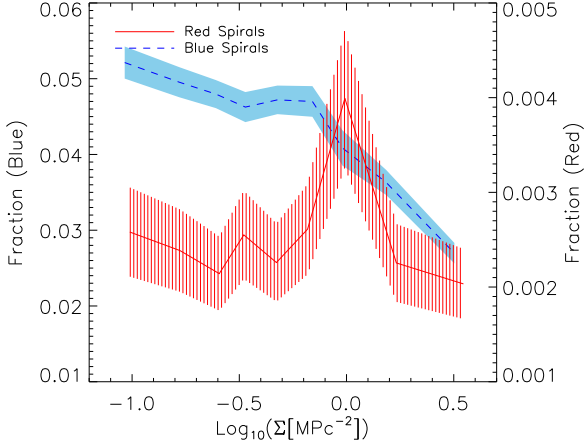}
\caption{Fraction of red spiral galaxies (red solid line) and blue spirals (blue dashed line) relative to all types of SDSS galaxies in our volume limited sample plotted versus environment as measured by $\Sigma$. The error regions are also shown; note that the binning is the same for both red and blue spiral samples, and adjusted to have $\sim30$ red spirals per bin. 
\label{4}}
\end{figure}

\begin{figure}
\includegraphics[width=84mm]{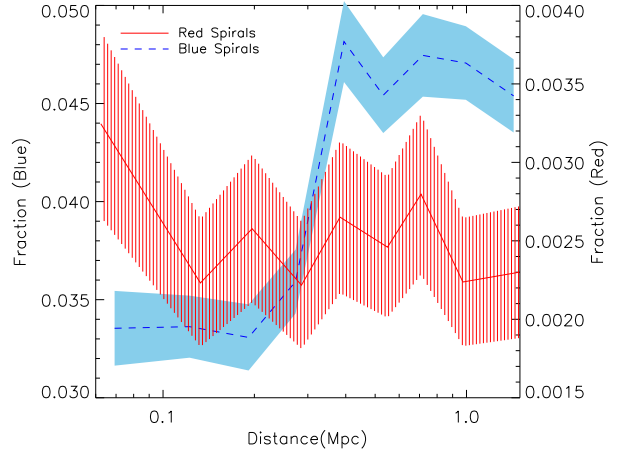}
\caption{Fraction of red spiral galaxies (red solid line) and blue spirals (blue dashed line) relative to all types of SDSS galaxies in our volume limited sample as a function of projected distance to nearest neighbour.  The error regions are also shown; note that the binning is the same for both red and blue spiral samples, and adjusted to have $\sim30$ red spirals per bin. \label{5}}
\end{figure}

In Figure \ref{4} we present the fraction of galaxies in our red and blue spiral samples (relative to the full volume limited galaxy sample) versus the local galaxy density. We confirm the previous findings that red spiral fraction increases with local galaxy density, along with the well established observation that the fraction of blue (or normal) spirals decreases. This figure suggests that the red spiral fraction peaks at intermediate densities, around $\Sigma = 1$ Mpc$^{-2}$, inside the infall radius of a typical cluster, but not in their cores (see Figure 5 of \citealt{B09} for the correlation of $\Sigma$ with distance to the nearest cluster centre, this level of $\Sigma$ corresponds to $d\sim5-15$ Mpc depending on the richness of the cluster - for example it could be quite close in to a poor group, or much further away from a rich cluster). In order to look at the impact of close neighbours, and not just local density, we also consider the (projected) distance to the nearest  neighbour (within 100 kms$^{-1}$ in the SDSS DR6 Spectroscopic Galaxy Sample). Figure \ref{5} shows the fraction of red and blue spirals in our face-on disky spiral sample as a function of this distance, showing that red spirals are a lot more likely to have close neighbours than blue spirals. (Of course $\Sigma$ and the distance to the nearest neighbour are correlated - galaxies with a higher $\Sigma$ will naturally have a nearer closest neighbour, and we have not attempted to disentangle the two measurements here.)

\subsection{Environmental Dependence of Star Formation}
\begin{figure}
\includegraphics[width=84mm]{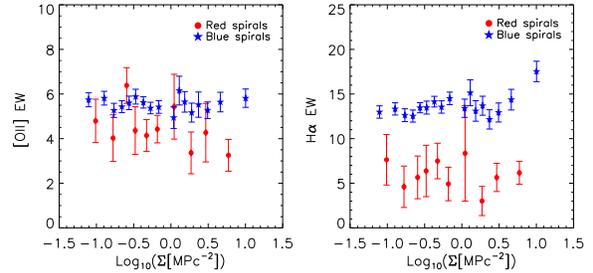}
\caption{The median equivalent widths of [OII] and H$\alpha$ as a function of local galaxy density.  Blue and red spirals are shown as blue stars/red dots respectively.  Error bars show our estimate of the $1\sigma$ error on the median value. 
\label{OIIIHa}}
\end{figure}

 We now study the environmental dependence of star formation tracers in our samples of red and blue spirals. In Figure \ref{OIIIHa} we show the median equivalent widths (EW) of [OII] and H$\alpha$ as a function of local galaxy density for both blue and red face-on disky spirals. 
  
  At all densities, red spirals have significantly lower but {\it non-zero} equivalent widths in these two tracers of ongoing star formation, but interestingly there are no significant trends with local density in either population, neither do we see any evidence for a change in the distribution of the quantities with environment (consistent with the findings of \citealt{Bam08} for H$\alpha$ in the galaxy population as a whole). There is perhaps a hint that the [OII] EW decreases with density for red spirals - a straight line fit to this trend has a $\sim2\sigma$ significance (a slope of $-1.0\pm0.4$). We also see a slight increase in the H$\alpha$ EW in blue spirals at very high density, which may be related to interaction triggered star formation in this population - or perhaps a higher rate of AGN contamination in blue spirals in high density regions. This figure includes all galaxies, including those classified as AGN by their optical spectra (see Section 5 below), however in Section 5 we will show there is a higher AGN fraction in the red spirals than the blue, so removing AGN contamination from these lines will only make the difference between red and blue spirals stronger. 
 
\begin{figure}
\includegraphics[width=84mm]{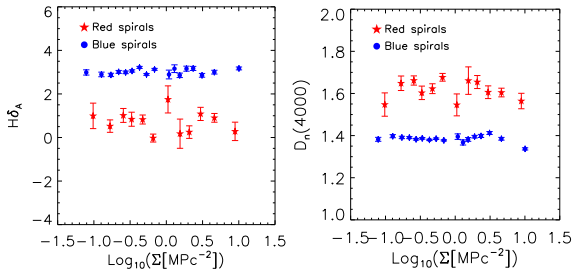}
\caption{Median values of H$\delta_A$ (a tracer of recent star formation) and $D_n(4000)$ (a tracer of average stellar age) as a function of local galaxy density.  Blue and red spirals are shown as blue stars/red dots respectively.  Error bars show our estimate of the $1\sigma$ error on the median value. 
\label{delta}}
\end{figure}

 Figure \ref{delta} shows similar relations to that shown in Figure 10 for the tracers H$\delta_A$ and $D_n(4000)$ (as discussed in Section 3, these are tracers of recent starbursts and the mean age of the stellar population respectively). Again we see that {\it at all densities} these indicators show less recent star formation and older stellar populations in red spirals than blue spirals, and that there are no significant trends with density in either population. We have repeated this exercise using (projected) distance to the nearest neighbour, and find similar results.
 
 Red spirals in all environments have lower rates of recent and on-going star formation than blue spirals. This suggests that despite the increased fractions of red spirals in high density regions, {\it environment is not the only factor in the shutting down of star formation in red spirals} - \ie~ the main physical process cannot only happen in high density regions (ruling out ram pressure stripping as the dominant process) and neither can it proceed faster in higher densities.  
 
 Our data are consistent with the process that turns spirals red being more likely to happen in high density regions as long as it is possible and proceeds in the same fashion in all environments. Both gentle interactions, and/or strangulation mechanisms could explain these observations.

\section{The Impact of AGN}

The most recent semi-analytic models for galaxy formation all invoke some form of AGN feedback which inhibits star formation in the most massive, red (early-type) galaxies \citep{G04,S05,Cr06,Bower06} in order to match the number counts and properties of these objects in the local universe. Observational evidence now exists to support this hypothesis, showing that the AGN fraction peaks in the region between the blue cloud and the red sequence \citep{S07,ScL09}. Could AGN feedback be responsible for the shutdown of star formation in the red spirals?

We test this by using emission line diagnostic diagrams (Baldwin Phillips \& Terlevich 1981, hereafter BPT; \citealt{VO87,Ke01,Ke06,CF01}) to probe the dominant source of ionisation in the red and blue spiral galaxies. For this study, we restrict our analysis to galaxies with greater than 2$\sigma$ emission line detection for all four lines used in the BPT diagram (i.e. [OIII], H$\alpha$, [NII] and H$\beta$). Likewise, in order to reduce the effect of aperture bias of the SDSS $3\arcsec$ fibre on the BPT diagram, we shift the lower redshift range of our sample up from $z>0.03$ to $z>0.05$. This reduces the sample size to 181 (60\%) of the red spirals, and 3462 (66\%) of the blue spirals. 

Standard emission line diagnostic diagrams \citep[e.g.][]{K03b,S07} are used to divide galaxies whose budget of ionising photons is dominated by current star formation (i.e. ionisation from OB associations) and those whose inter-stellar medium is excited by other processes. These other processes include AGN, but also shocks and evolved stellar populations such as post-asymptotic giant branch (pAGB) stars, extreme horizontal branch stars, white dwarfs \citep{St08,Sarzi09}. 

Observationally, the area between these two regions is populated by transition or composite objects where the total contribution to the ionising budget from star formation and other processes is roughly comparable. \citet{Ke01} give a prescription based on theoretical modelling of starburst galaxies which indicates the region of the diagram which can be explained by starburst emission (to the lower left of the line), while \citet{K03b} made an empirical fit to the observed separation between the populations in SDSS galaxies, which is often used as a lower limit for the location of AGN. Here as a short-hand, we will call objects to the upper right of the \citet{Ke01} line ``AGN" (even if other processes may be more important in some objects), or sometimes "Seyfert+LINER" (as explained below), objects below the \citet{K03b} line ``star-forming" and objects falling between the two lines ``composite" objects. 

The region dominated by ionization from processes other than star formation is furthermore divided into Seyferts and LINERs (low-ionisation nuclear emission-line regions; Heckman 1980). To split this region into Seyfert and LINER types we use the diagonal dividing line suggested by \citet{S07}. Objects in the Seyfert region are clearly identified as classical obscured (Type 2) Seyferts (\ie~ AGNs), while the LINER region is more controversial and possibly represents a heterogeneous population (see Section 6 of \citealt{H08} for a recent review of possible sources of ionisation for LINERs). Recent results from the SAURON survey by \citet{Sarzi09} show that the extended LINER emission seen in most SDSS fibre spectra (whose physical footprint corresponds to 2-3 kpc in the sample studied here) is inconsistent with a central point source for the ionisation, ruling out nuclear activity as the dominant source for the emission. Intriguingly, this is even the case when the presence of nuclear activity (e.g. from radio data) is detected.

\begin{figure*}
\includegraphics{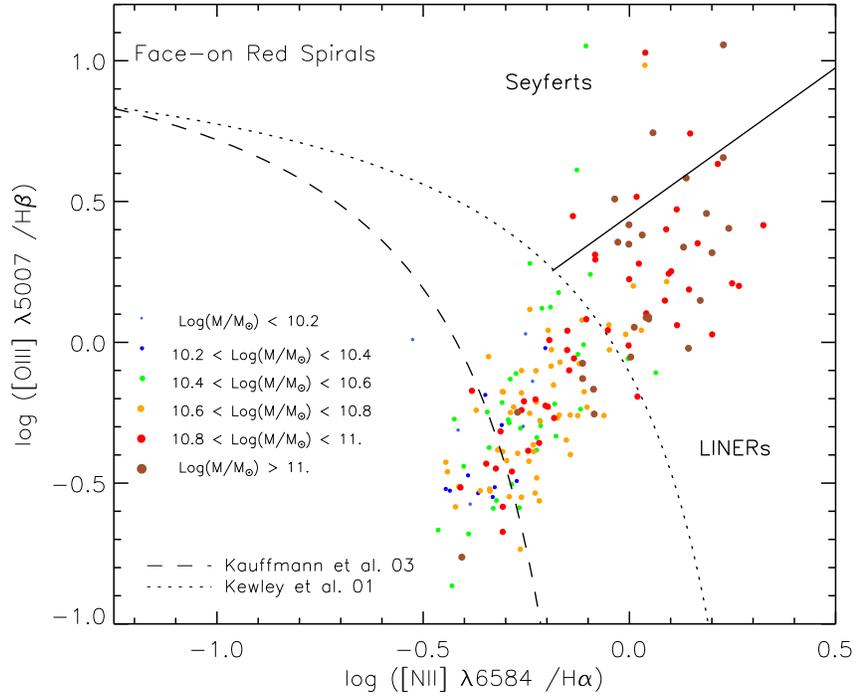}
\caption{Distribution of face-on disky red spiral galaxies in BPT diagram. The size and colour of the points indicates the stellar mass of the galaxies (as labelled). The diagonal dividing line for LINERs/Seyferts is taken from \citet{S07}.
\label{redBPT}}
\end{figure*}

The position of red spirals on a BPT diagram is illustrated in Figure \ref{redBPT} where points are coded (both in size and colour) by the stellar mass of the galaxy. Many more of the red spirals than the blue are placed above the \citet{Ke01} dividing line, indicating that they are not dominated by emission from star forming regions, but their gas must be ionised by other mechanisms ($30\pm4\%$\footnote{Errors on the fractions given in this Section and elsewhere are $\sqrt N$ counting errors. This obviously breaks down when fractions approach zero or one, however for the purposes of this work the approximation of the errors is adequate.}  of all face-on disky red spirals in redshift interval of 0.05-0.085, compared to $4\pm1\%$ of blue spirals). As the plot indicates, a significant fraction of these galaxies are LINER-type ($82\pm12\%$). The remaining red spirals have many more composite objects than are found in the normal spiral population ($49\pm5\%$ compared to $15\pm1\%$ of blue spirals) with only a relatively small fraction being classed as starforming by the \citet{K03b} criterion ($21\pm3\%$ of red spirals compared to $81\pm2\%$ of blue spirals). 

\begin{figure*}
\includegraphics{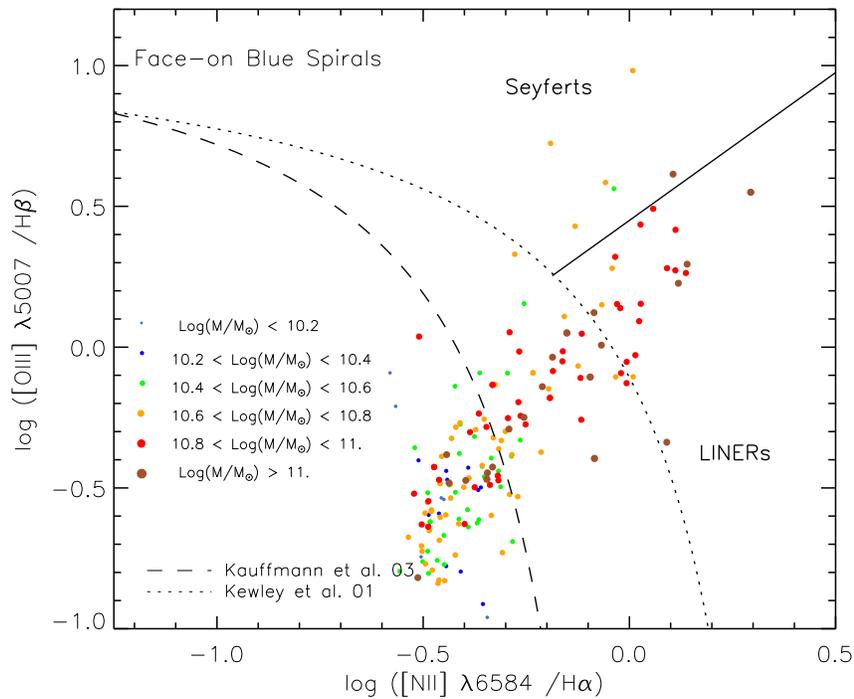}
\caption{Distribution on the BPT diagram of our sample of face-on disky blue spiral galaxies with the same number of objects in each mass bin as the red spirals shown in Figure \ref{redBPT} (selected randomly). The size and colour of the points indicates the stellar mass of the galaxies (as labelled).
\label{blueBPTmatched}}
\end{figure*}

 Part of these trends can be explained by the larger stellar masses of the red spirals, since it is well known that Seyferts and LINERs are more common in higher mass galaxies \citep[e.g.][]{K03b}. Therefore for a fair comparison we construct a sample of blue spirals selected from the full blue spiral population in such a way that they have the same mass distribution as the red spirals (our ``mass matched" blue spiral sample). We show 181 of these galaxies (\ie ~the same number as in the red spiral sample for ease of comparison) on the BPT diagram in Figure \ref{blueBPTmatched}. Compared to this mass matched sample of blue spirals, red spirals are still significantly more likely to host optically identified Seyfert+LINER emission lines. Recall, $30\pm4\%$ of red spirals are above the \citet{Ke01} division and $48\pm5\%$ are in the composite region. In comparison we find only $7\pm1\%$ mass matched blue spirals in the Seyfert+LINER region and $23\pm1\%$ in the composite region. Red spirals are therefore $\sim$4 times as likely to have Seyfert+LINER emission lines, and twice as likely to be composite objects when compared to similar blue spirals.
 
\begin{figure}
\includegraphics[width=84mm]{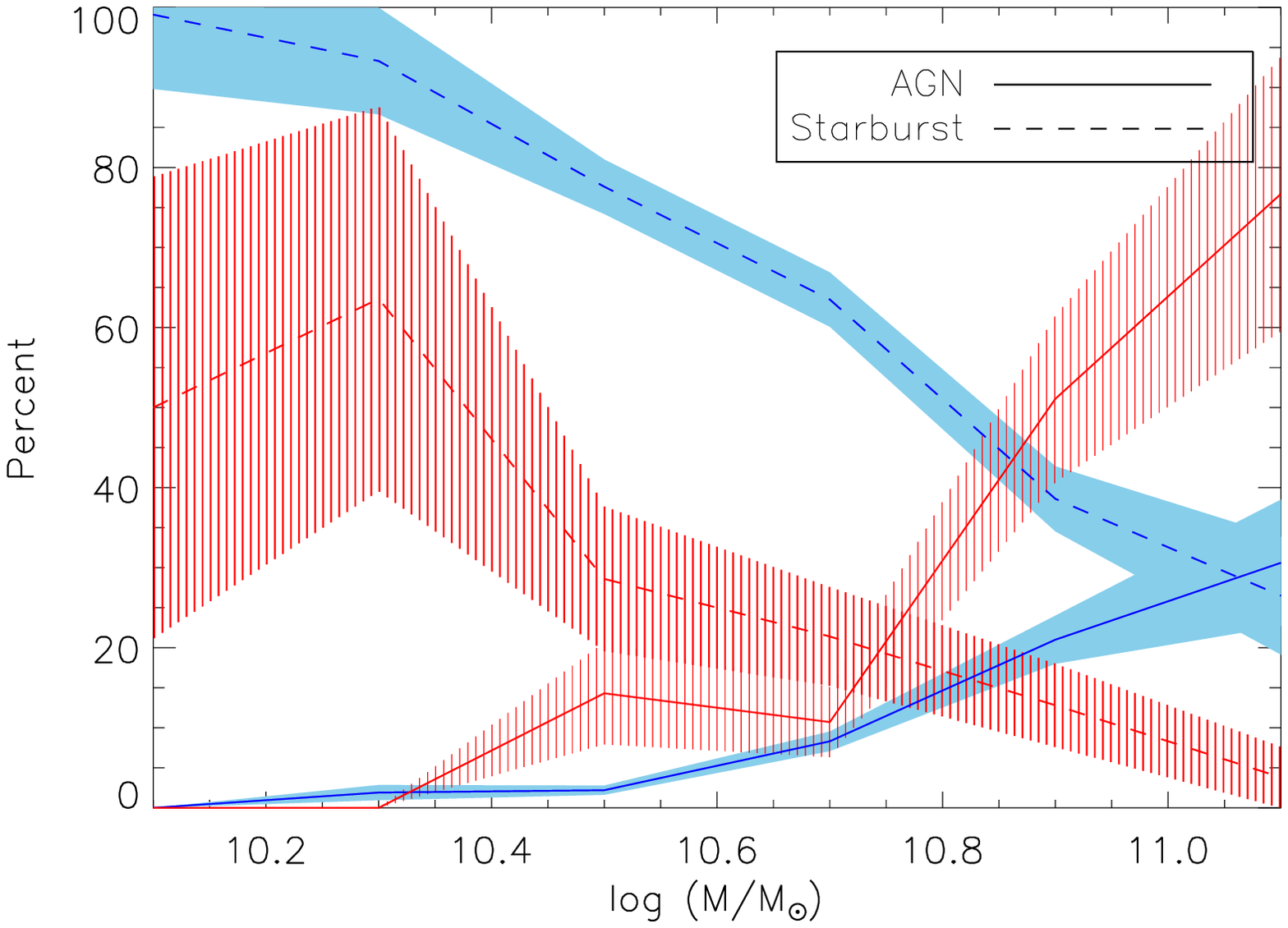}
\caption{Fraction of galaxies in a given stellar mass range which are classifed as "AGN" (Seyfert+LINER) or starforming from the BPT diagram (expressed as a percentage; note that the fraction classified as composites are not shown). The trends for blue spirals are shown with the blue lines, red spirals by the red lines. Shaded regions show the $\sqrt{N}$ counting errors on the fractions. Obviously these errors break down where the fractions approach zero or one, but give a reasonable estimate elsewhere. \label{bptfractions}}
\end{figure}

 We plot in Figure \ref{bptfractions} the observed fractions of Seyfert+LINER and starforming galaxies as a function of mass for both the red and blue spirals. For both sub-populations the Seyfert+LINER fraction increases with stellar mass, while the starforming fraction decreases, however as previously commented, at a given stellar mass red spirals are roughly twice as likely to be classifed as Seyfert/LINER, and less likely (roughly half as likely) to be classified as a starforming. There is a suggestion that the increase of Seyfert+LINER fraction with mass is also faster in the red spirals than it is in the blue spirals (at the expense of ``composite" objects, since the starforming fractions are seen to decrease at approximately the same rate).

 AGN fractions in all galaxies have generally been shown to be independent of environment (Miller et al. 2003, Sorrentino, Radovich \& Rifatto 2006). However, recent work on X-ray selected AGNs does suggest an environmental dependence such that X-ray selected AGN are more common in groups than clusters \citep[][this work also comments on the complete disjoint of X-ray selected and BPT selected AGN samples making such studies hard to compare]{Ar09} and \citet{L09} also suggest AGN are less common in higher densities. Here, we see no signature of a density dependence of the Seyfert+LINER fraction of red spirals, although we note that the red spiral sample is quite small to divide in this way and covers only a limited range of densities.

  Red spirals with emission lines which are not classified as coming from star formation are more likely to be classified as LINERS than similar objects in the blue spirals population, with $82\pm12\%$ of those in red spirals being LINERS versus $57\pm7\%$ in the mass matched blue spirals. This means the Seyfert fractions of red and blue spirals of the same mass are actually quite similar ($6\pm2\%$ of red spirals versus $3\pm1\%$ of blue spirals) - the main difference in the two populations appears to be in the LINER fraction which suggests a link between LINER emission and the shutting down of star formation in spiral galaxies. Overall $25\pm4\%$ of the red spirals are classified as having LINERs, while only $4\pm1\%$ of the (mass matched) blue spirals meet the LINER criteria. Interestingly in visually classified early type galaxies, \citet{S07} also found that an increase in LINER fraction was associated with ``stellar quiescence", while \citet{Sm09} studying radio loud AGN also found that the radio AGN in the red sequence were dominated by low ionization AGN which were classified as LINERs in the BPT diagram. 

\begin{figure}
\includegraphics[width=84mm]{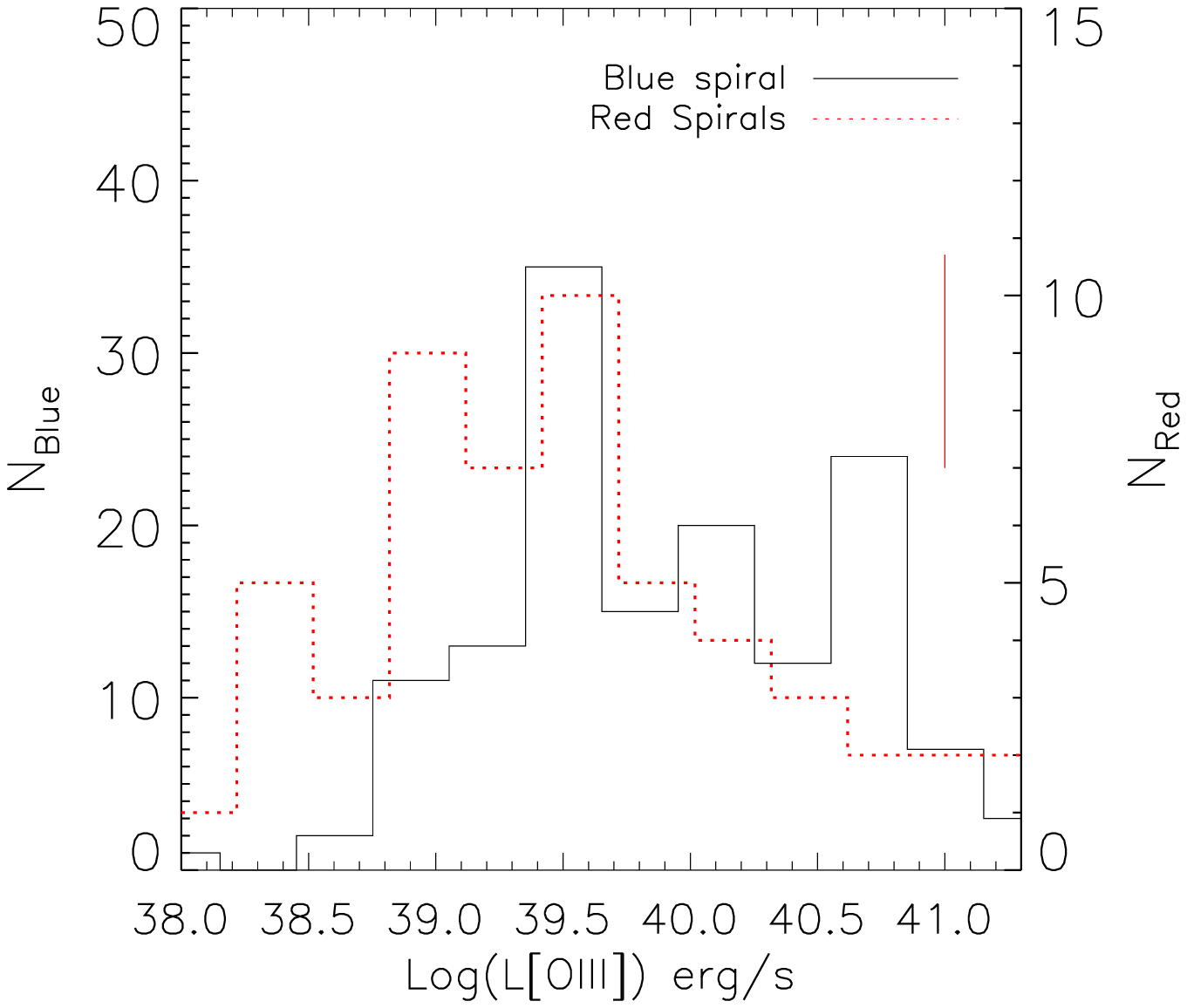}
\caption{Histogram of [OIII] luminosities of optically identified Seyferts and LINERS in the red spirals (red dotted line) and blue spirals from the mass matched sample (black solid line).  The typical error on the red spiral histogram is indicated, offset for illustration purposes. Note that the $y$-axes are scaled such that the total area under each histogram is equal. 
\label{LOIII}}
\end{figure}

 The Balmer line emission from even a small amount of star formation is expected to dominate any AGN emission. \citet{Sc09}  study the masking of AGN (Seyfert) emission in star forming galaxies and show that for AGN [OIII] luminosities below $10^{39}$ erg~s$^{-1}$ most galaxies with star formation present will not be classified as AGN, and that only above $10^{40}$ erg~s$^{-1}$ will the sample of AGN identified by optical emission lines be reasonably complete in the blue cloud. We show in Figure \ref{LOIII} histograms of the [OIII] luminosities (extinction corrected using the method described in \citealt{L09}) of both red and blue spirals (from the sample matched in mass with the red spirals) classified as having either Seyfert or LINER emission. This diagram suggests that there may be a difference in the luminosities of the emission lines identified in the two populations, such that fainter lines (below about $10^{39.3}$ erg~s$^{-1}$) are found in the red spiral population which are not found in the blue spiral population. However, since the actual number of red spirals are small, the histograms do agree within the error.
  
 If true, the above observation suggests that at least some LINERs are being revealed in the red spiral population, either because the emission lines are no longer dominated by star formation, or perhaps because the smaller amount of recent star formation means there is less dust shrouding the AGN.  However the difference cannot account for all of the extra LINERs in the red spirals - \ie~ even at the luminous end of the distribution where we would not expect LINERs to be hidden in the blue spirals, we find more LINERs in red spirals (considering only emission with $L$[OIII]$>10^{39}$ erg~s$^{-1}$ we find $24\pm4\%$ of the red spirals have emission not consistent with star formation compared to $7\pm1\%$ of the mass matched blue spirals), so we are able to say that at least high luminosity LINER emission is correlated with the shutting down of star formation in spiral galaxies.
 
Since we do not have either integral field observations or X-ray/radio data of red spirals, we present two alternative interpretations for the high LINER fraction in red spirals:

\begin{enumerate}
\item{\it Photoionisation from evolved stellar populations.}\\
If we assume that the LINER emission in red spirals is mostly due to photoionisation from evolved stellar populations \citep[e.g.][]{St08}, then we can naturally account for the relatively higher fraction of LINERs in red compared to blue spirals in the following way: in massive galaxies, there is a substantial component of hard ionising radiation coming from evolved stellar populations, that would, in the absence of other factors, produce a weak LINER spectrum. In blue spiral galaxies, this signature is mostly overwhelmed by ongoing star formation, whereas in red spirals, where star formation has been gently quenched, no such competing source of photons is available.

\item{\it Low-luminosity nuclear activity.}\\
If we assume that the LINER emission in red spirals is due to photoionisation from a low-luminosity, low-Eddington AGN (e.g. Kewley et al. 2006), we can ask about the role of this AGN in the evolution of red spirals. As suggested earlier, red spirals gently quenched their star formation in an extended process that began about 2 Gyr in the past. This in turn implies that the {\it current} LINER AGN can not be responsible for the suppression of star formation as it is happening now - with a substantial time delay after the quenching process. A similar, though shorter time delay is seen between the suppression of star formation and the start of substantial black hole growth (Schawinski et al. 2007). The LINER phase in red spirals, like in early-types (Schawinski et al. 2007), seems to be a post-quenching phenomenon. The interesting question then becomes why the particular internal environments of red spiral galaxies that have already quenched their star formation is so conducive to low-level black hole growth.
\end{enumerate}

Further observations of red spirals are required to fully distinguish between these two interpretations. Spatially resolved spectroscopic observations  (long-slit or IFU) are needed to test whether the LINER emission in red spirals is spatially extended as it is in early-types. Deep X-ray or radio observations could yield evidence for ongoing black hole accretion.

\section{Bar Fractions in Red/Blue Spirals}

 Even a cursory inspection of a small sample of images suggests that red spirals may have a significantly higher bar fraction than blue spirals (e.g. Figure 3). Two of us (BC and KLM) visually inspected the entire face-on disky red spiral sample, along with a similar number of the random mass matched blue spirals. We find that red spirals have an optical bar fraction of at least $67\pm5\%$ (raw fractions were $72\%$ for BC and $67\%$ for KLM) and up to 100\% including more uncertain identifications, while the blue spirals have a bar fraction of only $27\pm5\%$ (28\% from BC and 26\% from KLM). These bar identifications used the SDSS $gri$ images typically used by Galaxy Zoo, and were based on classic visual bar identification methods such as those used by \citet{RC3} to find a bar fraction of 25-30\% in the RC3. More recent work on bar fractions in the literature \citep[e.g.][]{J04,KS08,Bar08,A09} use automated techniques for finding bars using elliptical isophote fitting. The two studies using SDSS data on $\sim$2000 ``disk" galaxies \citep{Bar08,A09} are most directly comparable to this study. Both however use automated techniques to identify disk/spiral samples based largely on concentration \citep{A09} and colour \citep[][this study does also consider Sersic fits, but the final results are from colour selected ``spirals" only]{Bar08}, so we expect systematic differences with our visually identified face-on spiral sample and inparticular point out that red spirals by our definition will be completely absent from the \citet{Bar08} study, and extremely rare in \citet{A09}. The overall bar fractions found by \citet{Bar08,A09} are comparable to our total bar fraction of $\sim$50\% (they find 50\% and 45\% of their disk samples hosting bars), but intriguingly both studies suggest a trend for bluer disk galaxies to be more likely to host bars, in direct contrast to the clear signal we find for red spirals to have more obvious bars in the $gri$ images. It is not clear at this point if the difference in these trends is due to the sample selection or the bar identification method (although \citet{A09} claim only a 7\% difference between their automated bar finder and a visual check of their sample). A more detailed study of bar fractions in Galaxy Zoo galaxies as a function of various galaxy properties and considering possible biases on the visual bar identification method with $gri$ images is being prepared using bar identifications for almost 30,000 spiral galaxies collected during the second phase of Galaxy Zoo (GZ2; Masters et al. in prep.). For the purposes of this work, the huge increase in bar fraction  between the blue and red spiral samples gives such a strong hint of a trend of bar fraction with colour (in the sense that red spirals are much more likely to host bars), that it suggests bars may be providing an important clue to the formation of the red spirals and therefore the impact of bars should be considered in a discussion of their possible origin.
  
  In simulations of spiral galaxy formation, bars form quickly once a stable disk is formed, and are difficult to destroy \citep[e.g.][]{D06}. However, the impact of higher density environments on bars is unclear. Tidal interactions might induce bar formation (eg. Hernquist \& Mihos 1995), but they also act to heat the disks of spirals, and bars form most quickly in cold disks (eg. Toomre 1964). If external triggers such as tidal interactions are the most important source of bar instabilities, then the higher density environments of the red spirals may naturally lead to high bar fractions. 
  
  A possible explanation for the high bar fraction in red spirals could be that the bars themselves are at least partly responsible for the process which turned off star formation in these objects. Bars are known to be the most efficient way to redistribute material in the disks of galaxies (eg. Combes \& Sanders 1981), by channelling gas into the central regions of the galaxy. Bars have been invoked (eg. Shlosman et al. 2000) as a way to feed gas to the central black hole - the increase in LINER fraction we observe in the red spirals could be a remnant of this process (if they are LINERs associated with AGN). Perhaps the bars in the red spirals have removed the cold gas from the disk and channelled it inwards where it has either been used as fuel for the AGN, or created a starburst. Of course, this starburst must have happened more than $\sim1$ Gyr ago, to be consistent with our observation that red spirals are not post starburst objects, but if bars are as robust as simulations suggest then they would still persist for a long time after evidence of any bar triggered central starburst was removed.
   
\section{Discussion}
  
  Since red spirals are observed at all density levels, the process which creates red spirals cannot be confined only to regions of high galaxy density; environment alone is not sufficient to determine whether a galaxy will become a red spiral or not.  The lack of any clear correlation of the star formation rates of red spirals with environment, also indicates that red spirals in low density regions are similar to those in clusters making it more difficult to invoke only environmental processes in their formation. The process which turns spirals red may be more likely to occur in higher density regions, but must be possible, {\it and proceed in a similar way} in all environments, unless completely different mechanisms are responsible for red spirals in high and low density regions. That the red spiral population has a significantly higher mean mass than the population of blue spirals, also indicates that they are not uniformly sourced from the field population. However, perhaps the mass transition between red and blue spirals represents the lowest mass spiral which can retain its spiral structure under the influence of environmental effects which shut off its star formation (\ie. low mass blue spirals might pass through the red spiral phase very quickly, or experience both morphological and colour transformation at the same time).  
  
  We suggest that rather than representing an intermediate stage of environmental transformation, red spirals could be ``old spirals" who through normal internal evolution have already used up their reservoirs of gas - perhaps aided by the redistribution of gas due to a bar instability. We suggest that part of the reason they are found to be more common in higher density regions is because the initial density fluctuations decoupled from the Hubble flow earlier there \citep[see][for a review]{VB09} and galaxies started assembling at much earlier times \citep{C09} in dense environments so have had longer to use up their gas. Such objects would naturally be found at the high mass end of the galaxy distribution as they have been assembling for a long time. Red spirals presumably become less common in the highest densities where strong environmental processes are more important and may have already changed all spirals into early type galaxies.
  
  It has been shown that for a typical spiral (\ie~ with typical star formation rates) to use up all its gas within a Hubble time requires that there is no infall of additional gas from its halo \citep{L80}. Therefore strangulation (or removal of gas from the outer halo -  a process which would be more common in satellite galaxies in higher density regions, but can happen even in low mass groups \citealt{KM08}) is probably required and could by itself create a population of red spirals preferentially found in intermediate density regions (if those in the highest densities were further disrupted). It was observed by \citet{Sk09} that the Galaxy Zoo red spirals are preferentially satellite galaxies in massive halos (while blue spirals are preferentially the central galaxy in a low mass halo). This observation supports the idea of strangulation as an important part of the mechanism which creates red spirals. In this scenario they are accreted onto the massive halo as a blue spiral, after which their outer halo gas reservoirs are gently stripped by the main galaxy, and no further accretion of cold gas can take place.  \citep{Bekki02} showed that in such a scenario spiral arms would persist for a few Gyrs after the gas was removed. Lower mass spirals presumably are disrupted by gravitational effects in the massive halo and cease to display spiral features, so this model can explain both the mass and environmental distribution of the red spirals.
    
  If other environmental processes are responsible for turning spirals red, they must be very gentle and happen over long timescales. Continued tidal interactions and/or minor mergers (\ie ~galaxy harrassment, \citealt{M99}) could be responsible - these would heat the disk gas above the density required for star formation, but not remove it. If tidal interactions are also responsible for creating bars (e.g. Hernquist \& Mihos 1995) the high bar fraction we observe could be a remnant of this (although the bars may still help to shut down star formation by re-distributing and heating the gas themselves).
  
  Several authors \citep[e.g.][]{G03,W09} have suggested that red (or passive) spirals are the progenitors of S0 galaxies. The origin of the S0 galaxies has been an ongoing debate for more than three decades \citep{D80,M07}. S0s are often argued to be too bright and massive and have too large bulge-disk ratios to simply be faded spirals \citep[e.g.][]{B05}, but \citet{M07} argue that they have linked passive spirals with young S0s in two $z\sim 0.5$ clusters and that a combination of gas stripping and gentle galaxy-galaxy interactions could be responsible for the change, and \citet{D07} show that bulge light is severely attenuated by dust in spiral galaxies which may help to resolve the bulge-disk ratio issue. 
  
   There is only one observation in this paper which is immediately in conflict with the idea of the Galaxy Zoo red spirals being the progenitors of S0s. We observe a high optical bar fraction in the face-on disky red spirals, while S0s have a significantly lower bar fraction \citep{A09}. Our disky red spirals cannot simply turn into S0s as their spiral arms fade, as the bar instability would remain much longer than the arms. However in this work we have purposely selected red spirals with small bulges. Perhaps S0s form from red spirals with large bulges. Further study of the red Galaxy Zoo spirals with large bulges may clear up this issue. 
     
\subsection{Future Directions}     
     
 This work has used optical data from SDSS to reveal the star formation history, dust content, AGN properties, environments and morphologies of a unique sample of intrinsically red disk dominated spiral galaxies. It is clear that significantly more information can be obtained from other surveys, wavelengths, and from future theoretical modelling, which will help to shed light on the dominant physical process which formed these objects.  
  
 If strangulation is the dominant process forming red spirals they should have almost no residual neutral hydrogen (alternatively, if tidal interactions are responsible by gently heating the disk gas, and thus lowering its density below the threshold for star formation, they should still contain significant quantities of HI). This will be tested using data from the ALFALFA blind HI survey \citep{G05} which covers much of the SDSS footprint. 
 
 The debate over the nature of the LINER emission could be solved by moving to other wavelengths to reveal the AGN or by obtaining spatially resolved optical spectra to look at the extent of the emission. Deep targeted X-ray observations would reveal objects currently growing their black holes. Unfortunately, currently available public X-ray surveys are too small an area or at too low luminosity levels to be of use. (For example the XMM Cluster Survey, XCS \citep{XCS} has only 220 of the blue spirals and 6 of the red spirals studied here in its footprint  (1 blue spiral is detected), while the ROSAT All-Sky Survey, RASS \citep{RASS} has a high flux limit). Current radio continuum surveys (for example the NRAO-VLA Sky Survey, NVSS; \citealt{NVSS}) could also reveal ongoing black hole accretion, or radio-loud AGN. 
 
  More sophisticated modelling of the spectral information (for example from VESPA, \citealt{T09}) would provide time resolved star formation histories and more reliable stellar masses. Spatially resolved spectra would additionally allow separate study of the star formation histories of the different regions of these galaxies, and provide dynamical information possibly helpful in determining the role of the bars. 
   
 The interpretation of the increased bar fraction in the red spirals is made difficult by the wide range of suggestions of the impact of bars on spiral galaxies. More detailed modelling of spiral galaxies with bars would be extremely helpful in interpreting those observations. We can also use Galaxy Zoo 2 data (which asks for a more detailed classification of among other things barred galaxies) to study the more general properties of a large sample of galaxies with bars. For example studying the environmental dependence of barred galaxies compared to non-barred galaxies would shed light on the role tidal interactions play in bar formation. 
          
\section{Summary and Conclusions}
 We study the interesting population of red, or passive, spiral galaxies found by the Galaxy Zoo project. We identify from these red spirals a population of intrinsically red true disk dominated spirals by limiting the sample in inclination (to reduce the impact of dust reddening), requiring that spiral arms be visible, and removing bulge dominated systems using the SDSS parameter {\tt fracdeV}. We compare this sample to blue spirals selected in the same way and find that:
 
 \begin{itemize}
 \item{Red spirals are more likely to be more massive, luminous galaxies than blue spirals. They represent an insignificant fraction of the spiral population at masses below $10^{10}$\msun ~but are significant at the highest masses, showing that {\it massive galaxies are red regardless of morphology}.}
  
  \item{At the same mass as blue spirals, face-on red spirals do not have larger amounts of dust reddening (as measured by the Balmer decrement), therefore their red colours indicate an ageing stellar population not an increased dust content.}
  
  \item{Red spirals have lower (but not zero) rates of ongoing and recent star formation when compared to blue spirals. This is partly related to their higher average mass, however {\it at a fixed mass, red spirals still have less recent star formation than blue spirals}.}
  
\item{As previously observed, red spirals are more common at intermediate local densities (around, or just inside the infall regions of clusters). They are also observed to be more likely than blue spirals to have close neighbours.}

\item{Red spirals in all environments have lower rates of recent and on-going star formation than blue spirals, and there are no significant trends of the star formation rates with environment when spirals are split into red/blue. Clearly the process which creates red spirals is not confined to regions of high galaxy density. So {\it environment alone is not sufficient to determine whether a galaxy will become a red spiral or not}.}

  \item{Red spirals are more than four times more likely to be classified as Seyfert+LINER/composite objects from their optical spectra than blue spirals. This is partly due to the higher masses of red spirals but is still observed when they are compared to a blue spiral sample selected to have the same mass distribution. We find that a small fraction of low luminosity AGN are being revealed as the star formation is turned off in the red spirals, but this is not enough to account for all the difference. Most of the difference comes from an increased fraction of LINER-like emission ($82\pm12\%$ of Seyfert/LINERs found in red spirals are LINERs compared to $57\pm7\%$ in blue spirals).}
  
\item{Red spirals have significantly higher bar fractions than blue spirals (70\% versus 27\%), suggesting that bar instabilities and the shutting down of star formation in spirals are correlated}. 
\end{itemize}
 
  We propose three possible origins for the red spiral population studied in this work and suggest the most likely explanation is that a combination of the three accounts for the shutting down of their star formation while they retain their spiral structure:
      \begin{enumerate}
  \item{Perhaps red spirals are just old spirals which have used up all of their gas. They are found preferentially in intermediate density regions because structures first starts to form at the peaks of the dark matter distribution, but in the centres of clusters spiral morphologies cannot stand up to the environmental disturbances. Red spirals then represent the end stages of spiral evolution irrespective of environment (and in the absence of major mergers) - the spiral version of ``downsizing".}
\item{Perhaps red spirals are satellite galaxies in massive dark matter halos. In this scenario, they are accreted onto the halo as a normal blue spiral and have experienced either strangulation (where the gas in their outer halos has been gently stripped off, and no further cold gas has been allowed to accrete) or harassment (heating their disk gas and preventing further star formation). Low mass spirals would probably be disrupted in this process and so are not observed as red spirals.}
    \item{Perhaps red spirals evolved from normal blue spirals which had bars that were particularly efficient at driving gas inwards. This removed gas from the outer disk and turned the spiral red. If it triggered star formation in the central regions it must have occured more than $\sim 1$ Gyr ago since red spirals are not post starburst galaxies.}

  \end{enumerate}
  
  The red spirals in this work probably cannot be the progenitors of S0s as they have a significantly higher bar fraction than in observed in the S0 population. S0s may however be the end product of red spirals with larger bulges than we have studied here.

\paragraph*{ACKNOWLEDGEMENTS}

  This publication has been made possible by the participation of more than 160,000 volunteers in the Galaxy Zoo project. Their contributions are individually acknowledged at \texttt{http://www.galaxyzoo.org/Volunteers.aspx}.  KLM acknowledges funding from the Peter and Patricia Gruber Foundation as the 2008 Peter and Patricia Gruber Foundation International Astronomical Union Fellow, and from the University of Portsmouth and SEPnet (www.sepnet.ac.uk). Support for the work of MM in Leiden was provided by an Initial Training Network ELIXIR (EarLy unIverse eXploration with nIRspec), grant agreement PITN-GA-2008-214227 (from the European Commission). AKR, MM, HCC, RCN acknowledge financial support from STFC. Support for the work of KS was provided by NASA through Einstein Postdoctoral Fellowship grant number PF9-00069 issued by the Chandra X-ray Observatory Center, which is operated by the Smithsonian Astrophysical Observatory for and on behalf of NASA under contract NAS8-03060. CJL acknowledges support from The Leverhulme Trust and the STFC Science In Society Programme. Funding for the SDSS and SDSS-II has been provided by the Alfred P. Sloan Foundation, the Participating Institutions, the National Science Foundation, the U.S. Department of Energy, the National Aeronautics and Space Administration, the Japanese Monbukagakusho, the Max Planck Society, and the Higher Education Funding Council for England. The SDSS Web Site is http://www.sdss.org/.

\appendix
\section{Sample of Data Tables}

 We provide in this Appendix samples of the data tables we make available electronically listing important information (all based on SDSS data) for both our red (Table \ref{redtable}) and blue (Table \ref{bluetable}) face-on disky spiral samples. In both tables columns are (1) SDSS objid; (2 and 3) RA and Dec in J2000 decimal degrees; (4) Redshift; (5) Absolute $r$-band (Petrosian) magnitude, $M_r$; (6) $(g-r)$ colour (from model magnitues); (7) ``redness", defined as the distance in magnitudes from the blue limit of the $(g-r)$ verses $M_r$ red sequence (see Equation 1 in Section 2.1); (8) axial ratio, $\log (a/b)$ from the $r$-band isophotal measurement; and (9) {\tt fracdeV}; the fraction of the best fit light profile made up by a de Vaucouleur profile (as opposed to an exponential disk). 

\begin{table*}
\caption{A Sample of Face-On, ``Disky" Red Spirals. (The full table is available in the online version of the paper.)}
\label{redtable}
\begin{tabular}{lrrcccccc}
\hline
SDSS objid & RA (J2000 deg) & Dec (J2000 deg) & Redshift & $M_r$ & $(g-r)$ & ``Redness" & $\log (a/b) $ & {\tt fracdeV} \\
\hline
       587731186741346415 &   1.218164 &   0.351724 & 0.0833 & -21.26$\pm$0.01 &  0.695$\pm$0.020 &  0.040 &  0.144 & 0.42 \\
      588015509270364319 &   2.277073 &   0.042422 & 0.0735 & -21.75$\pm$0.01 &  0.701$\pm$0.021 &  0.036 &  0.143 & 0.46 \\
      587731187279265981 &   3.718539 &   0.634254 & 0.0642 & -21.15$\pm$0.01 &  0.756$\pm$0.039 &  0.103 &  0.042 & 0.00 \\
      587731187279331410 &   3.771462 &   0.642945 & 0.0621 & -21.50$\pm$0.01 &  0.662$\pm$0.017 &  0.002 &  0.179 & 0.32 \\
      588015508734148754 &   3.789492 &  -0.366536 & 0.0661 & -22.18$\pm$0.01 &  0.687$\pm$0.018 &  0.013 &  0.070 & 0.40 \\
  ... & ... & ... & ... & .. & ... & ... & ... & ...  \\

      \hline
\end{tabular}
\end{table*}
      
      \begin{table*}
\caption{A Sample of Face-On, ``Disky" Blue Spirals. (The full table is available in the online version of the paper.)}
\label{bluetable}
\begin{tabular}{lrrcccccc}
\hline
SDSS objid & RA (J2000 deg) & Dec (J2000 deg) & Redshift & $M_r$ & $(g-r)$ & ``Redness" & $\log (a/b) $ & {\tt fracdeV} \\
\hline
      587727225690259639 &   0.276560 & -10.400263 & 0.0754 & -21.41$\pm$0.01 &  0.434$\pm$0.020 & -0.225 &  0.084 & 0.12 \\
      587727225690259660 &   0.325174 & -10.426144 & 0.0767 & -21.03$\pm$0.02 &  0.561$\pm$0.028 & -0.090 &  0.078 & 0.07 \\
      587730773888991473 &   0.566363 &  14.671807 & 0.0834 & -20.91$\pm$0.02 &  0.503$\pm$0.035 & -0.145 &  0.106 & 0.06 \\
      588015510343385206 &   0.665200 &   0.942673 & 0.0807 & -22.38$\pm$0.01 &  0.632$\pm$0.018 & -0.046 &  0.079 & 0.22 \\
      588015510343516321 &   0.941711 &   0.854914 & 0.0611 & -20.69$\pm$0.01 &  0.378$\pm$0.023 & -0.266 &  0.185 & 0.00 \\
      ... & ... & ... & ... & .. & ... & ... & ... & ...  \\
      \hline
\end{tabular}
\end{table*}
      
      \end{document}